\documentclass[12pt]{article}
\usepackage{a4wide}
\usepackage{amssymb}
\usepackage{graphicx}
\graphicspath{{images/}}
\begin{document}
{\renewcommand{\thefootnote}{\fnsymbol{footnote}}
\begin{center}
{\LARGE  Adherence and violation of the equivalence principle from classical to quantum mechanics }\\
\vspace{1.5em}

Joseph Balsells\footnote{e-mail address: {\tt balsells@psu.edu}}
and Martin Bojowald\footnote{e-mail address: {\tt bojowald@psu.edu}}
\\
\vspace{0.5em}
Institute for Gravitation and the Cosmos,\\
The Pennsylvania State
University,\\
104 Davey Lab, University Park, PA 16802, USA\\
\vspace{1.5em}
\end{center}
}

\setcounter{footnote}{0}

\begin{abstract}
  Investigation into the applicability of the equivalence principle in quantum
  mechanics has taken many forms, with varying conclusions. Here, a dynamical
  semi-classical description of a wave packet in terms of its center of mass
  and higher quantum fluctuations is applied to the case of a quantum particle
  in gravitational free fall. The analysis provides an intuitive account of
  the origin of mass-dependence in quantum-gravitational dynamics through an
  effective potential that enforces the uncertainty principle. This potential
  has two implications: (i) The lowest order quantum fluctuations encoding the
  width and spreading of the wave packet obey an uncertainty relation whose
  observance is mass-dependent. (ii) In an inhomogeneous gravitational field
  tidal effects couple the center of mass motion to the quantum
  fluctuations. The combined effect results in a clear demonstration of how
  some conceptions of the weak equivalence principle, based on mass
  dependence, are violated. The size of this violation is within sensitivities
  of current E{\"o}tv{\"o}s and clock-based return time experiments.
\end{abstract}

\section{Introduction}

For over a century, general relativity has been a cornerstone for our
understanding of gravity. However, despite its remarkable success, general
relativity presents inconsistencies with quantum theory. A significant area of
conflict between the two theories has been their different predictions for the
mass-dependence of gravitational phenomena. The present study revisits
this conflict and clarifies how mass-dependence emerges as one incorporates
quantum effects into gravitational physics in the weak-field regime. At
  the same time, we will construct useful new methods that may be applied to
  derivations of quantum dynamics relevant for potential experiments in this context.

In general relativity, gravity manifests as the metric geometry of spacetime and particle trajectories are determined by the geodesics of this spacetime. In the weak-field limit of the theory, the geodesic equation is expressed in terms of the perturbation tensor \(h_{ab}\) as \cite{carroll}
\begin{equation}
  \label{eq:weak-field-geodesic}
  \frac{d^2x^i}{dt^2} = \frac{1}{2} \frac{\partial h_{00}}{\partial x_i}\,.
\end{equation}
This result yields a set of coupled ordinary differential equations for the coordinates \(x^i\) of a freely falling particle, determined without any information about the particle mass. The principle of weak equivalence elevates this model-dependent result to a general physical principle often stated as the independence of the future history of a particle in gravitational free fall from its specific properties. Tino et al. recently provided a comprehensive review \cite{tino2020} of the status of the equivalence principle and its tests, which presents compelling experimental support for the mass-independence of particles in gravitational free fall.

Quantum theory contests this claim. Indeed, quantum theory's necessary
dependence on mass is evident in the commutation relation
\([\hat{x},\hat{p}] = i \hbar\). The presence of the dimensionful quantity
\(\hbar\) in this relation makes it generally impossible to rescale the
equations of quantum theory in a way which eliminates the implicit
mass-dependence from any calculation. The manner in which quantum theory's
predictions either violate or adhere to the equivalence principle in specific
cases is increasingly well-understood theoretically. For example,
Greenberger's early study \cite{greenberger1968} demonstrates that applying
quantum theory to a particle bound in an external gravitational potential
leads to mass-dependent predictions for observables, including energy levels,
frequencies, and orbital radii. Mass-dependence has also been predicted for
physically more relevant dynamical wave packet states in \cite{viola1997}.

Following this analysis, Greenberger proposed the abandonment of the equivalence principle within quantum theory \cite{greenberger1968}. In support of this proposal, Sonego demonstrated how the weak equivalence can be dispensed with as a fundamental principle without compromising the formal apparatus of general relativity \cite{sonego1995}. Okon and Callender view these developments as indicating that the question of whether quantum phenomena adhere to or violate the equivalence principle has been resolved \cite{okon2011}. These authors contend that further quantum tests of the equivalence principle are unlikely to yield substantial theoretical insights. Yet, in embracing this standpoint, it remains essential to address the mechanism behind the vanishing of mass-related influences on gravitational phenomena in the classical limit---a challenge that the authors of this study contend has yet to be satisfactorily met.

For example, in Greenberger's study \cite{greenberger1968} the disappearance of mass in the classical limit is explained via a quantization condition where the quantum number characterizing the state becomes proportional to the particle mass in the classical limit (therefore cancelling it from all results) without explaining how this equivalence arises. This present study provides a new analysis which better clarifies how mass-dependence emerges as quantum effects are incorporated into gravitational physics in the weak-field regime.

The outline for the paper is the following. In Section \ref{sec:methods} we
review an uncommon geometric formulation of quantum mechanics. We discuss the
kinematical and dynamical aspects of the theory, its relation to the wave
function formulation, and highlight the role geometrical quantum theory plays
in providing a structure on which to define a consistent
semiclassical hierarchy. In Section \ref{sec:applications}, this framework is
applied to the problem of quantum particle motion in an external gravitational
field. We obtain equations of motion incorporating mass-dependent quantum effects
which permit us to address a number of issues surrounding the quantum theory of motion in a gravitational background field.

Section \ref{sec:eotvos} introduces the E{\"o}tv{\"o}s framework for parametrizing violations of the weak equivalence principle. We apply our analysis to compute non-zero quantum corrections to the E\"otv\"os parameter. The corrections we develop depend on wave packet contributions which typical E{\"o}tv{\"o}s experiments may not be sensitive to. However, as established by the foundational studies \cite{kasevich1991,kasevich1992,peters2001}, atom interferometers are well-suited to testing weak-field aspects of general relativity due to their precise control over test matter preparation and observation. Our analysis is particularly relevant to atom interferometric E\"otv\"os tests, as these experiments involve quantum matter manipulation. In Section \ref{sec:propagation-phase} we indicate how our methods can be used to obtain the interferometer phase, generalizing the methods used in \cite{hogan2008}.

Atom interferometric E{\"o}tv{\"o}s tests probe the particle only at the set of positions defined by the laser pulse sequence used. Limitations of this design are detailed in \cite{nobili2020}. Clock experiments provide a complementary framework for testing the equivalence principle with quantum matter. In these experiments one characterizes motion in a gravitational potential as barrier scattering and tracks the return time for a particle launched into the potential. If particles of differing masses and matching initial conditions are found to return in different times, this would signal a violation of the weak equivalence principle. Time of flight measurements depend on properties integrated across a particle's entire trajectory. These experiments may therefore be sensitive to violations in the weak equivalence principle not easily seen in E{\"o}tv{\"o}s experiments.

In Section \ref{sec:returntime} we address also the problem of geodesic motion of quantum particles from this point of view. We provide a new analysis of the return time of a quantum object thrown up in a gravitational field which benefits from the fully dynamical equations of motion for a quantum particle we develop here.

We conclude in Section \ref{sec:conclusion} with a brief discussion of the general features of our analysis which may prove useful in future studies.

\section{Canonical Effective Methods}
\label{sec:methods}

The mathematical structures underlying classical and quantum physics appear
very different, a fact which can complicate the understanding of conflicting
predictions like the mass-dependence of particle motion. However, this
difference can be better understood thanks to a geometric formulation of
quantum mechanics in which the classical limit may be carefully defined. Here
we review the elements of this theory only as they are relevant to the problem
posed in modeling a quantum particle in a gravitational field. Mathematically
precise treatments of the general theory may be found in the references
\cite{strocchi1966}, \cite{kibble1979}, \cite{ashtekar1999}, \cite{bojowald2006}, and also \cite{bjelakovic2005}.

\subsection{The space of states and observables}
\label{sec:space-stat-observ}

As starting point in modeling a quantum system, we make a choice of a unital
operator algebra \(\mathcal{A}\) specifying the relevant observables. In this
paper we choose the algebra generated by position and momentum operators
satisfying the canonical commutation relation
\begin{equation}\label{eq:ccr}
  \left[ \hat{x}, \hat{p} \right] = i \hbar.
\end{equation}
In the analytical description of quantum theory, we would next choose a
representation of this algebra by operators acting on a separable complex
Hilbert space \(\mathcal{H}\) and define states as positive trace-class linear
operators on \(\mathcal{H}\). For example, in this formulation one typically
denotes a pure state \(\rho\) in terms of a representative
\(\psi \in \mathcal{H}\) as
$\rho = \vert \psi \rangle \langle \psi \vert/\langle \psi \vert \psi
\rangle$. It is easy to show that a state thus defined is insensitive to
arbitrary complex (and possibly time dependent) rescaling of the
representative:
\begin{equation}\label{rescale}
  \vert \psi \rangle\mapsto f(t) \vert \psi \rangle
\end{equation}
for
\(f(t) \in \mathbb{C}\) at fixed $t$. This result indicates that we may equivalently
identify the pure states of a quantum system with the rays of
\(\mathcal{H}\). The collection of rays of \(\mathcal{H}\) form the projective
Hilbert space.

The geometric formulation arises as an alternative to the analytical formulation by taking seriously that the projective Hilbert space, and not \(\mathcal{H}\) itself, provides the correct space of states. As a subspace, the projective Hilbert space has the structure of a symplectic manifold called the quantum phase space \(\Gamma\) (for additional detail on this, see \cite{strocchi1966,kibble1979}). Two consequences of this characterization follow. First, the physical states contain all physical information about the system. Choice of a specific wave function state \(\psi \in \mathcal{H}\) from its projection onto \(\Gamma\) is non-unique and requires additional (non-physical) information. This point will be especially relevant in our discussion of the interferometer phase in Section \ref{sec:propagation-phase}. Second, the phase space characterization facilitates identifying the classical phase space as a sub-manifold of the quantum phase space.

In the geometric formulation of quantum mechanics, as in classical mechanics, a point \(p\in \Gamma\) specifies the state of the system. Observables are constructed as smooth real-valued functions on the quantum phase space \(F\colon\Gamma \rightarrow \mathbb{R}\).
A useful set of observables for the algebra generated by (\ref{eq:ccr}) is the set consisting of the action of the state on the algebra generators, \(   \left\langle    \hat{x}  \right\rangle\) and \(   \left\langle    \hat{p}  \right\rangle\), together with higher central moments of the state defined in a completely symmetric ordering as
\begin{equation} \label{eq:momentsdef}
  \Delta \left(x^{o-m}p^m\right)
  \equiv \big\langle ( \hat x - \langle\hat{x}\rangle )^{o-m} (\hat p -
  \langle\hat{p}\rangle)^m  \big\rangle_{{\rm symm}}
\end{equation}
where $o\geq 2$ and $0\leq m\leq o$ are integers.
The utility of these functions are several. For one, they summarize statistical information about the state and we refer to these observables as moments of the quantum state. For a given moment observable we will call the quantity \(o\) the moment's order.
Second, these observables are used to establish the semiclassical condition of a quantum state. We say a state is semiclassical if the moment observables evaluated on this state satisfy the hierarchy condition
\begin{equation}
  \label{eq:hierarchy}
  \Delta(x^ap^b) = O(\hbar^{(a+b)/2}).
\end{equation}
Such conditions are satisfied for Gaussian states, but also by more general states because the specific coefficients of $\hbar^{(a+b)/2}$ are not determined by the condition (\ref{eq:hierarchy}).

\subsection{Poisson structure and dynamics}
\label{sec:poiss-struct-dynam}

The symplectic structure on the state space of quantum mechanics can be
expressed through a Poisson bracket. A Poisson bracket acts on functions on a
symplectic or Poisson manifold, which we introduce by using arbitrary
operators $\hat{A}$ and $\hat{B}$ acting on the Hilbert space. The expectation
values $g_{\hat{A}}(\psi)=\langle\psi|\hat{A}|\psi\rangle$ and
$g_{\hat{B}}(\psi)=\langle\psi|\hat{B}|\psi\rangle$ can then be interpreted as
functions on the Hilbert space because they depend on the state $\psi$ in
which they are computed. The condition that $\psi$ be normalized implies
unique values on each ray of the Hilbert space, and we can view
$g_{\hat{A}}$ and $g_{\hat{B}}$ as functions on the projective Hilbert
space. The normalization condition does not completely eliminate the rescaling
freedom (\ref{rescale}), which is still possible by a phase factor $f(t)$ with
$|f(t)|^2=1$. However, the expectation value functions $g_{\hat{A}}$ and
$g_{\hat{B}}$ are independent of this remaining freedom. Therefore, they only
capture physical information about quantum states.

Given these functions on state space, their Poisson bracket is defined by
\begin{equation}
  \{g_{\hat{A}},g_{\hat{B}}\}
  =
  \frac{1}{i\hbar} \langle [ \hat{A}, \hat{B}] \rangle= \frac{1}{i\hbar}g_{[\hat{A},\hat{B}]}\,.
  \label{eq:1}
\end{equation}
This definition can directly be applied to powers and products of the basic
operators $\hat{x}$ and $\hat{p}$, and to moments (\ref{eq:momentsdef}) if we
use linearity and the Leibniz rule. Elementary discussion and applications of
this structure can be found in \cite{bojowald2022}. In particular, a
semiclassical truncation in which only moments up to a given order $a+b$ in
the hierarchy (\ref{eq:hierarchy}) are used leads to Poisson submanifolds that
are in general not symplectic. The Poisson tensor of the bracket
(\ref{eq:1}) restricted to such a subspace is then non-invertible, such
that there are so-called Casimir functions $C$ which have vanishing Poisson
brackets with all other functions in the same truncation.

The Poisson structure allows us to associate to each observable a vector field generating a Hamiltonian flow on the phase space. The dynamics are specified by the Hamiltonian vector field of a distinguished observable, the quantum Hamilton function obtained from the Hamiltonian operator \(\hat{H}\) for the quantum system as \(H = \langle \hat{H} \rangle\). Dynamics are obtained directly from the phase space structure:
\begin{equation}
  \frac{{\rm d}}{{\rm d}t} A = \{A,H\}.
\end{equation}
Based on (\ref{eq:1}), this dynamics is equivalent to the familiar Ehrenfest
theorem for the dynamics of expectation values.

By definition, the quantum Hamilton function $H=\langle\hat{H}\rangle$ is a function
on quantum phase space obtained by evaluating the expectation value of the
Hamiltonian operator in a generic state. When we parameterize states by
their basic expectation values $\langle\hat{x}\rangle$ and
$\langle\hat{p}\rangle$ together with the central moments, $H$ becomes a function
of these variables. A general expression for this function showing its dependence on these variables can be obtained
from a series expansion centered around the basic expectation values:
\begin{eqnarray}
    H
    &=& \left\langle H(\langle\hat{x}\rangle + (\hat{x} -
      \langle\hat{x}\rangle), \langle\hat{p}\rangle +
      (\hat{p}-\langle\hat{p}\rangle))\right\rangle \nonumber\\ 
    &=& H_{{\rm class}}(\langle\hat{x}\rangle,\langle\hat{p}\rangle)
    + \sum_{o=2}^\infty \sum_{m=0}^o \frac{1}{o!}{o\choose m}
    \frac{\partial ^{\,o}H}{ \partial x^{o-m} \partial p^{m}} \Delta \left(x^{o-m}p^m\right).
  \label{eq:Heffexpand}
\end{eqnarray}

The structure of the series expansion is revealing. First, we see that quantum dynamics reduce to their underlying classical analog when \(H\) can be expressed as a quadratic function of \(x\) and \(p\). This characteristic contributes to the prevalence of quadratic potentials in modeling quantum systems and explains the extensive research devoted to systems governed by quadratic potentials in the literature, for example in that of \cite{lammerzahl96}, where the Newtonian gravitational potential is approximated by its second order series expansion. Second, when dealing with non-quadratic potentials, the remaining terms in this expansion reveal the emergence of quantum effects. States whose fluctuations are non-zero are extended. In the higher-order terms, quantum fluctuations of the state couple to the external field through a derivative expansion of the potential. This coupling structure shows how the non-local nature of quantum dynamics appears for extended states in inhomogeneous fields.

In the non-quadratic setting the quantum effects can dominate or act as perturbations depending on the relative magnitude of higher-order terms. If the Hamiltonian remains polynomial, the series terminates at a finite order and perturbation theory may not be necessary.
Otherwise for non-polynomial interactions, we must consider the convergence properties of the
series (\ref{eq:Heffexpand}). In this case, systems satisfying the moment hierarchy condition, equation
(\ref{eq:hierarchy}), are well-behaved as higher-order terms are suppressed by powers of \(\hbar\). Truncating the expansion at a finite order yields a closed semiclassical dynamics with controlled errors. This dynamics agrees with the classical dynamics at zeroth order in \(\hbar\) but introduces perturbative couplings from higher moments. These perturbative effects may capture interesting quantum properties of the system as demonstrated in \cite{aragon-munoz2020} for the case of tunneling. In this work, we apply this method to analyze the
emergence of mass-dependence of trajectories in gravitational free-fall for
quantum systems.

\subsection{Canonical structure}
\label{sec:canonical-structure}

The first non-trivial quantum effects appear at second order in moments. Up to
second order there are two basic expectation values and three fluctuation moments. Their non-vanishing Poisson brackets are
\begin{eqnarray}
  \label{eq:2ndOrderBrackets}
    \{\langle\hat{x}\rangle,\langle\hat{p}\rangle\} &=& 1 \\
    \{ \Delta(x^2), \Delta(xp) \} &=& 2\Delta(x^2)\\
    \{ \Delta(xp), \Delta(p^2) \} &=& 2\Delta(p^2)\\
    \{ \Delta(x^2), \Delta(p^2) \} &=& 4\Delta(xp).
\end{eqnarray}
The odd dimension implies that the phase space is not symplectic.  Moreover,
the brackets are not canonical, but the Darboux theorem (or its generalization
to Poisson manifolds \cite{Weinstein}) guarantees that we can transform to canonical
coordinates. In this case, if we make the transformation
\begin{equation}
  \label{eq:canonicalVariables}
  \Delta(x^2) = s^2, \quad \Delta(xp) = s p_s, \quad \Delta(p^2) = p_s^2 + \frac{U}{s^2}
\end{equation}
then \(s\) is a configuration variable for the wave packet width and \(p_s\) its conjugate momentum such that
\begin{equation}
  \label{eq:6}
  \{s,p_s\} = 1.
\end{equation}
This transformation, without the background of Poisson geometry, has been
found several times independently in a variety of fields
\cite{VariationalEffAc,GaussianDyn,QHDTunneling}. A derivation from Poisson
geometry and generalizations to higher orders and two degrees of freedom can
be found in \cite{Bosonize,EffPotRealize}.

In (\ref{eq:canonicalVariables}), the variable \(U\), a Casimir function, is a
conserved quantity with dimensions of action squared satisfying
\begin{equation}
  \label{eq:uncertainty}
  \Delta(x^2)\Delta(p^2) - \Delta(xp)^2 = U.
\end{equation}
(Geometrically, hypersurfaces of constant $U$ in phase space are symplectic leaves of the Poisson manifold
that admit canonical coordinates $(x,p)$ and $(s,p_s)$.)  The
transformation to canonical variables therefore shows that
\(U\) is the phase space uncertainty volume for the wave packet, and the
second-order dynamics conserves its value. If higher-order moments are considered, we
would find that the product of second-order moments in
equation~(\ref{eq:uncertainty}) need not be conserved exactly, but still satisfy the
usual uncertainty inequality. Likewise higher order moments are subject also to uncertainty relations.
Higher order relations are developed for quantum states in \cite{brizuela2014}, although we will not need them here.

We choose to measure \(U\) in units of the minimum action squared
\begin{equation}
  U = \lambda U_{{\rm min}} = \lambda \frac{\hbar^2}{4}
\end{equation}
where \(\lambda\geq1 \) is dimensionless. The correct value of \(\lambda \) for a given problem will depend on the preparation of the state. To keep the calculation transparent and focus on the concepts, we consider the case \(\lambda  = 1\), which is correct for Gaussian states, as explained below.

\subsection{Generation of wave function states from moments}
\label{sec:state-reconstruction}

Canonical effective methods work directly with the observable quantum
statistics \(\langle\hat{x}\rangle,\langle\hat{p}\rangle,\) and \(\Delta(x^ap^a)\). These statistics may be measured and predicted independently of a specific choice of wave function state. Nonetheless, questions arise about whether these statistics encode all the physical information about the state, if there is redundancy in this choice of statistics, and how these statistics are related to the alternative description of quantum states using wave functions. In this section we provide a procedure for generating a wave function state \(\psi(x)\) from specified moment data.

\subsubsection{General considerations}
\label{sec:gener-cons}

Extracting moments from wave function states is straightforward. However, the
inverse task, constructing a wave function state compatible with specified moment data,
is more challenging. Indeed, in general neither existence nor uniqueness of such a state is guaranteed. The mathematical literature refers to the task of determining a distribution that generates a given set of moments as the problem of moments. A historical perspective on the moment problem, along with its extension into complex function theory, is presented in \cite{kjeldsen1993}.

In simple terms, we can ensure the existence of a real-valued distribution whose moments match a given sequence of numbers \(m_j\) by confirming that the Hankel matrices \((H_n)_{ij} = m_{i+j}, i+j\le n\) are positive definite for all \(n\in \mathbb{N}\). An accessible proof of this statement can be found in \cite{schmudgen2017}. However, in our subsequent application, we will assume that a wave function state exists based on physical reasoning, without examining the positivity of Hankel matrices constructed from the moments.

The uniqueness problem is nuanced. Non-uniqueness in the choice of a wave function state appears in two ways. First, a conventional wave function state encodes information for an infinite set of moments.
This allows for the existence of multiple non-identical states that share identical low-order statistics, making it impossible to reconstruct a unique state for our truncated moment system.
Second, even when selecting a wave function compatible with the provided moments, an additional freedom persists due to the complex rescaling (\ref{rescale}).
Consequently, we present a procedure for obtaining a specific state from the space of states compatible with the provided data. Our procedure generates a wave function state in the polar form
\begin{equation}
  \label{eq:polarDecomposition}
  \psi(x,t) = \sqrt{\rho(x,t)} \exp( i \theta(x,t))
\end{equation}
by first building the probability density \(\rho\) and then the phase, \(\theta\), out of moment data. The extension of this procedure to states described by density matrices is discussed in \cite{EffPotRealize}.

Results obtained through moment evolution and this procedure should agree with
experimental results, but may not agree with results obtained from the
Schr{\"o}dinger wave function theory. Such disagreements do not have
  physical implications because they merely correspond to different rescaling
  choices of the form (\ref{rescale}). Examples of disagreement with the
Schr{\"o}dinger theory are presented in the applications, Sections
\ref{sec:free-particle} and \ref{sec:evol-line-potent}. An agreement of this
method when used to determine the interferometer phase identified by certain
experiments is presented in Section \ref{sec:propagation-phase}.

\subsubsection{Density reconstruction}

If the unknown probability density \(\rho\) can be expressed as a
polynomial in \(x\), the reconstruction problem is linear and has a unique solution. However, due to normalization constraints, \(\rho\) typically is not polynomial. Nonetheless, the simplicity of reconstructing polynomials suggests a general approach: we decompose \(\rho\) into a polynomial basis that approximate it. We then reconstruct these approximations order-by-order to achieve the desired level of accuracy.

Following this idea, let \(L_n(x)\) be a complete, orthogonal set of polynomials with weight function \(w\) on \(L^2(w, \mathbb{R})\) and let \(u_n(x)\) be the associated orthonormal basis such that
\begin{equation}
  \label{eq:orthogonality}
  \int_\mathbb{R} u_n(x)u_k(x) dx = \delta_{nk}.
\end{equation}
with
\begin{equation}
  \label{eq:orthonormalfunctions}
  u_n(x) = \frac{1}{\sqrt{N_n}} \sqrt{w(x)}L_n(x).
\end{equation}
Having assumed the basis property of the \(u_n\), any function \(f(x)\) in \(L^2(w, \mathbb{R})\) can be expanded with coefficients in \(\mathbb{R}\) as
\begin{equation}
  \label{eq:expansion}
  f(x) = \sum_{n=0}^\infty c_nu_n(x)
\end{equation}
with coefficients
\begin{equation}
  \label{eq:coefficients}
  c_n = \int f(x) u_n(x) dx.
\end{equation}
In particular, we can reconstruct the density \(\rho\) from moment data if we choose \(f(x) = \rho(x)/\sqrt{w(x)}\). In this case, the expansion coefficients reduce expectation values of polynomials:
\begin{equation}
  \label{eq:34}
  c_n = \int \frac{\rho(x)}{\sqrt{w(x)}} \frac{1}{\sqrt{N_n}} \sqrt{w(x)}L_n(x) dx 
      = \frac{1}{\sqrt{N_n}} \langle L_n(\hat x) \rangle.
\end{equation}
Because \(L_n(x)\) is a polynomial in \(x\),  \(\langle L_n(\hat x)\rangle\) can be reconstructed from moments
\begin{equation}
  \label{eq:polynomialExpectation}
  \langle L_n\rangle = \sum_{k=0}^n l_{n,k} \langle \hat x^k \rangle.
\end{equation}
Expressing the expectation value in coefficient form uses the so-called raw forms of the moments, not the centralized ones. The two are nonetheless related by the binomial theorem
\begin{equation}
  \label{eq:39}
  \Delta(x^a) = \langle (\hat{x} - \langle\hat{x}\rangle)^a\rangle
  = \sum_{i=0}^a {a\choose i} (-1)^{a-i} \langle\hat{x}\rangle^{a-i} \langle\hat{x}^i \rangle
\end{equation}
which is a matrix equation that can be inverted to solve for the \(\langle\hat{x}^i\rangle\) from the provided \(\Delta(x^a)\).

Tracing these steps backwards gives finally the distribution reconstructed from its moments as
\begin{equation}
  \label{eq:reconstruction}
  \rho(x) = w(x) \sum_{n=0}^\infty \frac{1}{N_n}\langle L_n(\hat x)\rangle L_n(x)
  = w(x) \sum_{n=0}^\infty \sum_{j=0}^n \sum_{k=0}^n \frac{1}{N_n}
   l_{n,j} l_{n,k} \langle \hat x^j \rangle  x^k.
\end{equation}

\subsubsection{Phase reconstruction}

Moments of the form \(\langle \hat{x}^n\hat{p} \rangle\) can be used to
reconstruct the phase. These non-symmetric moments can be obtained as linear
combinations of symmetrically ordered ones. The real part of these moments
are given from the definition as
\begin{eqnarray}
  \label{eq:37}
  \mathfrak{R} \left(\langle \hat{x}^n\hat{p} \rangle\right)
  &=& \mathfrak{R} \int dx\, \psi^* \left(x^n \frac{\hbar}{i} \frac{d}{dx}\right) \psi\nonumber \\
  &=& \mathfrak{R} \int dx \sqrt{\rho}\exp(-i\theta) x^n \frac{\hbar}{i} \left[ \frac{d\sqrt{\rho}}{dx} \exp(i\theta)  + \sqrt{\rho} i\frac{d\theta}{dx} \exp(i\theta) \right]\nonumber\\
  &=& \hbar\int dx\, x^n \rho  \frac{d\theta}{dx}.
\end{eqnarray}
The function that multiplies the monomial powers of \(x\) in this is \(\hbar \rho d\theta/dx\). Consequently, we can apply the reconstruction procedure from before on this product with the outcome:
\begin{equation}
  \label{eq:generalPhaseDerivativeReconstruction}
  \frac{d\theta}{dx} = \frac{w(x)}{\hbar\rho(x)} \sum_{n=0}^\infty \sum_{j=0}^n \sum_{k=0}^n \frac{1}{N_n}
   l_{n,j} l_{n,k} \mathfrak{R}\left(\langle \hat{x}^j\hat{p} \rangle\right)  x^k.
\end{equation}
This result determines the phase derivative from moment data because the density's dependence on moments is already established.
Together with equation (\ref{eq:reconstruction}), these results provide the link between moments and wave function states.

The reconstruction procedure works by adapting a reference
distribution---the selected weighting function \(w(x)\)---to nearby
distributions such that the result matches the specified statistics. Here ``nearby'' means that only finitely many Taylor coefficients change. In a finite
truncation of the procedure there are many nearby distributions having the
same statistics depending on the arbitrary choice of orthogonal polynomial system. We focus
our examples on the useful choice of generalized Hermite polynomials which are
characterized by the shifted and rescaled Hermite weight function
\(\exp(-\frac{(x-m)^2}{2\alpha})\).
This choice allows us to encompass not only Gaussian states but also provides a structured approach for handling states that go beyond the Gaussian approximation.

Explicitly substituting generalized Hermite polynomials into the reconstruction provides the first order in moments approximations
\begin{eqnarray}
  \rho(x; m,\alpha)
  &=&
  \frac{1}{\sqrt{\pi \alpha^2} }e^{- \frac{\left(x - m\right)^2}{\alpha^2}}
  \left(
    1 + 2 \frac{m^{2}}{\alpha^{2}}  - \frac{2 m \langle \hat x \rangle}{\alpha^{2}} - 2 \frac{m x}{\alpha^{2}} + 2 \frac{\langle \hat x \rangle x}{\alpha^{2}}
    \right)
    \\
  \frac{d\theta}{dx}(x; m,\alpha)
  &=& \frac{\langle \hat p \rangle}
  {\hbar \left(1 + 2 \frac{m^{2}}{\alpha^{2}} - \frac{2 m \langle \hat x \rangle}{\alpha^{2}} - \frac{2 m x}{\alpha^{2}} + \frac{2 \langle \hat x \rangle x}{\alpha^{2}}\right)}.  
\end{eqnarray}
Centering the generalized Hermite functions about the center of mass with the choice \(m = \langle \hat x \rangle\) simplifies these expressions to
\begin{eqnarray}
  \rho(x)
  &=&
    \frac{e^{- \frac{\left(x - \langle \hat x\rangle\right)^2}{\alpha^2}}}{\sqrt{\pi \alpha^2} }
    \label{eq:firstOrderReconstructionRho}
  \\
  \frac{d\theta}{dx}
  &=& \frac{\langle \hat{p} \rangle}{\hbar}
    \label{eq:firstOrderReconstructionTheta}
\end{eqnarray}
where \(\alpha\) is still arbitrary because we have not assumed any second order statistics.

\begin{figure}
    \centering
    \includegraphics[width=6cm]{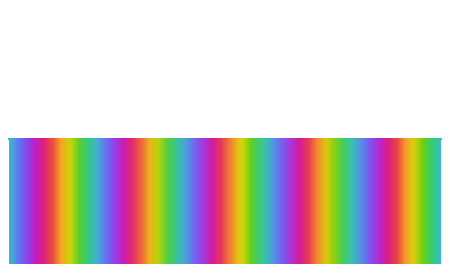}
    \caption{An abs-arg plot of a plane wave demonstrating its linear phase profile.
    \label{fig:planeWave}}
\end{figure}

For fixed choices of the first order data at a time \(t\), the phase derivative may be integrated with respect to \(x\) to give the phase profile of the instantaneous state as
\begin{equation}
  \label{eq:19}
  \theta(x,t)
  = \theta(x_0) + \frac{\langle \hat{p} \rangle(t) ( x - x_0) }{\hbar}.
\end{equation}
This linear phase profile matches that of a plane wave with momentum
\(\langle \hat{p} \rangle\) so we refer to the first order result as the plane
wave approximation to the phase, see Figure \ref{fig:planeWave}. In this approximation, the phase at any position is known if the phase at any other position and the (mean) momentum of the state are known.

In Section \ref{sec:propagation-phase}, this first-order reconstruction will be shown already to reproduce the interferometer phase identified for plane-wave states evolving in linear and quadratic potentials as presented in \cite{storey1994} and \cite{peters1998}. We extend those results by carrying the reconstruction to the next order (i.e. first non-trivial quantum order). Incorporating second order quantum fluctuations and choosing \(\alpha^2 = \Delta(x^2)\) provides the reconstructions
\begin{eqnarray}
  \rho(x)
  &=&
    \frac{ e^{-\frac{(x-\langle \hat x \rangle)^{2}}{2\Delta( x^2 )}}}
    {\sqrt{2\pi\Delta(x^2)}}
  \label{eq:secondOrderReconstructionRho}  \\
  \frac{d\theta}{dx}
  &=&
    \frac{\langle \hat p \rangle}{\hbar}
    + (x- \langle \hat x \rangle)
    \frac{ \Delta(xp) }
    {\hbar \Delta(x^2)}\,.
    \label{eq:secondOrderReconstructionTheta}  
\end{eqnarray}

\begin{figure}
    \centering
    \includegraphics[width=6cm]{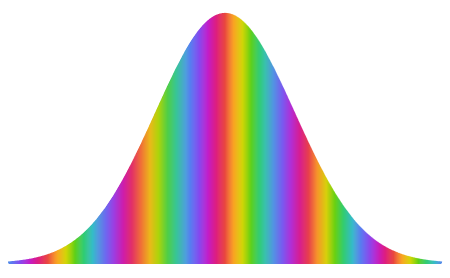}
    \caption{A Gaussian state as defined in equation
      (\ref{eq:freeReconstruction}). The parameters for this state were chosen
      to give a narrow momentum distribution about the same mean momentum as
      the plane wave in Fig.~\ref{fig:planeWave}. 
      \label{fig:narrowGaussian}}
\end{figure}

\begin{figure}
    \centering
    \includegraphics[width=6cm]{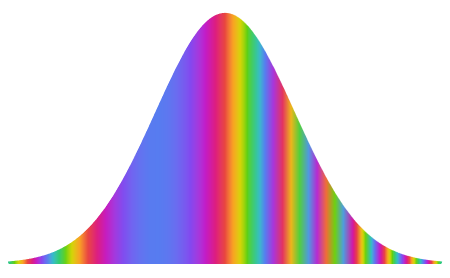}
    \caption{A Gaussian state with the same mean momentum but with a wide
      momentum distribution compared with Fig.~\ref{fig:narrowGaussian}.
    \label{fig:wideGaussian}}
\end{figure}

The probability density obtained in this case reproduces the well-known formula for a Gaussian probability density parametrized by its first two statistical moments. More interestingly, the phase derivative gains an additional term which is non-zero for \(x\ne\langle \hat{x} \rangle\). When looked at nearby to the wave packet center, the phase of a Gaussian state resembles that of a plane wave with momentum \(\langle \hat{p} \rangle\) and has a well-defined wavelength, see Figures \ref{fig:narrowGaussian} and \ref{fig:wideGaussian}. Corrections to the plane wave phase due to spatial localization become important when displacements from the wave packet center are significantly larger than the ratio of second-order moments \(\Delta(xp) / \hbar \Delta(x^2)\).

The spatial dependence of corrections is better understood when the phase derivative formula is expressed in canonical coordinates (\ref{eq:canonicalVariables}). In this form, the phase derivative is given by:
\begin{equation}
  \label{eq:secondOrderReconstructionThetaCan}
  \frac{d\theta}{dx} =
  \frac{ p }{\hbar}
  +
  \frac{x- \langle \hat x \rangle}{s}
  \frac{ p_s }  {\hbar}
\end{equation}
with \(s=\sqrt{\Delta(x^2)}\) giving the standard width of the packet. This result resembles that of a plane wave when either \(p_s\) is small, indicating narrow momentum spread, or when looking near the distribution center where \((x- \langle \hat x \rangle)/s < 1\).

Figure \ref{fig:narrowGaussian} illustrates the case of a state with narrow momentum spread. Its phase closely matches that of a plane wave for most of its weight. Conversely, when the momentum distribution is wide (and \(p_s>0\)), the phase increases more rapidly than a plane wave for \(x>\langle\hat{x}\rangle\), resulting in wavelength compression, and more slowly for \(x<\langle\hat{x}\rangle\), causing wavelength elongation. These corrections to the plane wave result are evident in Figure \ref{fig:wideGaussian} where parameters are chosen for a Gaussian wave packet with a broad momentum distribution.

For fixed choices of statistics, the equation (\ref{eq:secondOrderReconstructionTheta}) (or its canonical form, equation (\ref{eq:secondOrderReconstructionThetaCan})) may be integrated and combined with the density reconstruction, equation (\ref{eq:secondOrderReconstructionRho}), to give the instantaneous pure state reconstruction
\begin{equation}
  \label{eq:freeReconstruction}
  \psi(x) =
  \left(
    \frac{ 1 }
    {2\pi\Delta(x^2)}
  \right)^{1/4}
  \exp\left[
    -\frac{(x-\langle \hat x \rangle)^{2}}{4\Delta( x^2 )}
    \left(
      1
      -
      i \frac{ \Delta(xp) }
      {\hbar/2 }
    \right)
    + i \frac{\langle \hat p \rangle x }{\hbar}
    + i\theta_0
  \right].
\end{equation}
The overall phase \(\exp{(i\theta_0)}\) incorporates both the arbitrary
integration constant from equation (\ref{eq:secondOrderReconstructionTheta})
as well as the unspecified choice of branch from the square root of
\(\rho\). It is interesting to compare this result with the general form for a Gaussian pure state.

In the position basis, a general Gaussian wave packet can be represented as
\begin{equation}
  \label{eq:generalGaussian}
  \psi(x; A,B,C) = \exp{
    \left(
      A x^2 + Bx + C
    \right)}
\end{equation}
with \(A,B,C \in \mathbb{C}\). We identify the real degrees of freedom as
\begin{eqnarray}
  A &=& -(a + i \alpha) \\
  B &=& b + i \beta \\
  C &=& c + i \gamma
\end{eqnarray}
and require \(a>0\) so that the state is normalizable.
A tedious, but straightforward process of evaluating expectation value integrals and solving systems of equations identifies these parameters in terms of the state's statistical moments as
\begin{eqnarray}
  \label{eq:44}
  a &=& \frac{1}{4\Delta(x^2)} \\
  b &=& \frac{\langle \hat x \rangle}{2 \Delta(x^2)} \\
  \alpha &=& -\frac{\Delta(xp)}{2\Delta(x^2) \hbar} \\
  \beta &=& \frac{\langle \hat p \rangle}{\hbar} - \frac{\Delta(xp) \langle \hat x \rangle}{\Delta(x^2)\hbar} \\
  c &=& -\frac{b^2}{4 a} + \log
  \left[
    \left(\frac{2a}{\pi }\right)^{\frac{1}{4}}
\right]
\end{eqnarray}
with the overall phase \(\gamma\) undetermined.
With this choice of parametrization, the general Gaussian wave packet written in (\ref{eq:generalGaussian}) is expressed in terms of its statistics as
\begin{equation}
  \label{eq:18}
  \psi(x) =
  \left(  \frac{1}{2\pi \Delta(x^2)} \right)^{1/4}
  \exp \left[
    -\frac{(x-\langle \hat x \rangle)^2}{4\Delta(x^2)} \left(1 - i \frac{2\Delta(xp)}{\hbar} \right)
    + i \frac{\langle \hat p \rangle x}{\hbar}
    + i \gamma
  \right].
\end{equation}
Up to an overall (\(x\)-independent) phase, this result is identical to equation (\ref{eq:freeReconstruction}). Agreement between these two results indicates that our reconstruction procedure with the choice of generalized Hermite polynomials contains Gaussian states. When higher-order fluctuations are provided, the general reconstruction equations (\ref{eq:reconstruction}) and (\ref{eq:generalPhaseDerivativeReconstruction}) allow state approximations beyond the Gaussian form to be derived.

\section{Applications}
\label{sec:applications}

Our quasiclassical model can be used to address several questions related to quantum test masses travelling in a gravitational background field. We motivate our approach through the Ehrenfest equation. This equation relates the acceleration of the barycenter of a particle in a non-local quantum state to the potential energy \(V\) as
\begin{equation} \label{eq:ehrenfest}
  m\frac{d^2}{dt^2}\langle \hat{x} \rangle
  = - \left\langle \frac{d V}{d x} (\hat{x}) \right\rangle.
\end{equation}
For a Newtonian gravitational potential \(V(x) = m\Phi(x)\), with \(\Phi(x)\) independent of mass the Ehrenfest equation reduces to an equation without explicit mass-dependence:
\begin{equation} \label{eq:ehrenfest2}
  \frac{d^2}{dt^2}\langle \hat{x} \rangle
  = - \left\langle \frac{d\Phi}{d x} (\hat{x}) \right\rangle.
\end{equation}
In the classical theory, we identify the metric component \(h_{00}\) from the geodesic equation (\ref{eq:weak-field-geodesic}) with the Newtonian gravitational potential as
\begin{equation}
  \label{eq:13}
  h_{00} = - 2 \Phi.
\end{equation}
This leads to the equation of motion for a classical test mass
\begin{equation}
  \label{eq:weak-field-geodesic-2}
  \frac{d^2x}{dt^2} = - \frac{d \Phi}{d x}\,.
\end{equation}
 
Similar to its classical counterpart, the absence of explicit mass-dependence in the quantum equation hints at an extension of the classical weak equivalence principle into quantum theory. L{\"a}mmerzahl proposes such a quantum equivalence principle in \cite{lammerzahl96}. However, the absence of explicit mass-dependence conceals a subtlety. Unlike the classical equation, the quantum result forms a closed system of differential equations only when \(\Phi(x)\) is at most quadratic in \(x\). 
In general, equation~(\ref{eq:ehrenfest2}) must be supplemented with equations for higher moments (or suitable closure conditions and truncations that parameterize the values of higher moments).

We provide these additional equations from the effective Hamiltonian function associated with the Newtonian potential. Limiting the expansion (\ref{eq:Heffexpand}) to second order in moments provides the lowest-order quantum corrections
\begin{equation}
  \label{eq:Heff2}
  H_{{\rm eff},2}\left(x,p, \Delta(x^2), \Delta(xp),\Delta(p^2)\right)
  =
  \frac{p^2}{2m} + m \Phi(x)
  + \frac{\Delta(p)^2}{2m}
  + \frac{1}{2} m
  \frac{d^{2}\Phi}{ dx^{2}} \Delta(x^{2}).
\end{equation}
The coupling between quantum and classical degrees of freedom in the final term depends on the gravitational field curvature given by \(d^{2}\Phi/ dx^{2}\) which, being the second derivative of the metric, can be considered part of the Riemann tensor. 

We present a case analysis to emphasize the role of this coupling. In Section \ref{sec:free-particle}, we examine the scenario with \(\Phi = 0\), representing a free particle. In Section \ref{sec:evol-line-potent}, we select \(\Phi = g x\) to represent a particle in a linear gravitational potential. These cases illustrate classical center-of-mass dynamics with decoupled quantum dynamics. Subsequently, we explore quadratic and higher-order cases, discussing their implications for equivalence principle violation in the E{\"o}tv{\"o}s framework (Section \ref{sec:eotvos}) and the return time framework (Section \ref{sec:returntime}). Finally, in Section \ref{sec:propagation-phase}, we apply these methods to determine the interferometer phase shift in a Mach-Zehnder atom interferometer.

\subsection{Free particle}
\label{sec:free-particle}

An extensive investigation of the free particle case, \(\Phi = 0\), using the geometric point of view implied by our methods appears in \cite{bojowald2022}.
Using the brackets for second order moments, (\ref{eq:2ndOrderBrackets}), and the second-order effective Hamilton function (\ref{eq:Heff2}) with \(\Phi = 0\) produces the Hamilton equations of motion for the quantum free particle
\begin{eqnarray}
  \frac{d\langle \hat{x} \rangle}{dt} &=& \frac{\langle\hat{p}\rangle}{m} \\
  \frac{d\langle\hat{p}\rangle}{dt} &=& 0 \\
  \frac{d\Delta (x^2)}{dt} &=& \frac{2 \Delta(xp)}{m} \\
  \frac{d\Delta (xp)}{dt} &=& \frac{\Delta (p^2)}{m} \\
  \frac{d\Delta (p^2)}{dt} &=& 0.
\end{eqnarray}
These equations have solutions in terms of initial data as
\begin{eqnarray}
  \label{eq:freeSolutions}
  \langle\hat{x}\rangle(t) &=& x_0 + \frac{p_0}{m}(t-t_0) \\
  \langle\hat{p}\rangle(t) &=& p_0 \\
  \Delta (x^2)(t) &=& \Delta (x^2)_0 + \frac{2 \Delta(xp)_0}{m} (t-t_0) + \frac{\Delta (p^2)_0}{m^2}(t-t_0)^2 \\
  \Delta (xp)(t) &=& \Delta (xp)_0  + \frac{\Delta (p^2)_0}{m}(t-t_0)\\
  \Delta (p^2)(t) &=& \Delta (p^2)_0 
\end{eqnarray}
The first two of these equations indicate that for a localized free particle state, the center of mass trajectory is the classical one. The remaining equations of motion allow us to infer quantum mechanical effects.

To lighten notation we will assume that the initial conditions are specified such that
\(t_0 = 0\) when the wave packet is minimally squeezed, that is,
\(\Delta(xp)_0 = 0\). Additionally, we define shorthand for the initial width
and spreading frequency via
\begin{eqnarray}
  \sigma^2 &=& \Delta(x^2)_0\\
  \omega_\sigma^2 &=& \frac{\Delta (p^2)_0}{m^2\Delta (x^2)_0}.
\end{eqnarray}
With these choices and substituting the solutions (\ref{eq:freeSolutions}) into the Gaussian template equation~(\ref{eq:freeReconstruction}), we reconstruct the wave function 
\begin{equation}
  \label{eq:freeReconstructed}
  \psi(x,t) =
  \left(  \frac{1}{2\pi \sigma^2\left(1 + \omega_\sigma^2 t^2\right)} \right)^{\frac{1}{4}}
  \exp \left[
  - \frac{(x-x_0 - p_0t/m)^2}{4\sigma^2\left(1 + \omega_\sigma^2 t^2\right)}
  \left(1 - i \omega_\sigma t \right)
  + i \frac{p_0 x}{\hbar}
  + i \gamma
  \right]
\end{equation}
with the unspecified phase \(\gamma\).

Although it has been argued that obtaining the specific wave function solving the Schr{\"o}dinger equation is rarely necessary, the moment formalism provides an approach to doing so in which the partial differential equation is replaced by a system of ordinary differential equations which may be easier to solve. We see how this works out in this case. Substituting the result (\ref{eq:freeReconstructed}) into the time-dependent free particle Schr{\"o}dinger equation reveals a differential equation for \(\gamma(t)\):
\begin{equation}
  \label{eq:gammaEqn}
  \frac{d \gamma}{dt} = -\frac{p_0^2}{2 m \hbar } - \frac{1}{2}\frac{\omega_\sigma}{ 1+\omega_\sigma^2t^2}.
\end{equation}
This separable equation shows that in order for our reconstruction to solve the Schr{\"o}dinger equation, we must choose the \(x-\)independent phase to satisfy
\begin{equation}
  \label{eq:8}
  \exp(i\gamma)
  = \exp
  \left(
    - i \frac{p_0^2}{2m\hbar} t
    -\frac{i}{2} \arctan(\omega_\sigma t)
  \right).
\end{equation}
With this, we have demonstrated that the reconstruction (\ref{eq:freeReconstructed}) occupies the same ray of the Hilbert space as the wave function solving the Schr{\"o}dinger equation.

\subsection{Evolution in a linear potential}
\label{sec:evol-line-potent}

In a linear potential \(V(x) = m g x\), the equations of motion for the second-order statistics are identical to those for the free particle, but the solutions for the packet's centroid now satisfy
\begin{eqnarray}
  \langle\hat{x}\rangle(t) &=& x_0 + \frac{p_0}{m}t - \frac{1}{2} gt^2 \\
  \langle\hat{p}\rangle(t) &=& p_0 - mgt .
\end{eqnarray}
The construction of a falling Gaussian wave packet is immediate from the reconstruction template, equation~(\ref{eq:freeReconstruction})
\begin{eqnarray}
  \psi(x,t)
  &=&
    \left(  \frac{1}{2\pi \sigma^2\left(1 + \omega_\sigma^2 t^2\right)} \right)^{\frac{1}{4}}
    \\
  &&\times
    \exp
    \left[
    - \frac{\left(x-x_0 - \frac{p_0}{m} t + \frac{1}{2} gt^2\right)^2}
    {4\sigma^2\left(1 + \omega_\sigma^2 t^2\right)}
    \left(1 - i \omega_\sigma t \right)
    +
    i \frac{\left( p_0 - mgt \right) x}{\hbar}
    +
    \frac{i}{\hbar} \frac{g t^2}{2}
    \left(
    p_0 - \frac{ mgt}{3}
    \right)
    \right].\nonumber
\end{eqnarray}

As a point of comparison, the construction of a falling Gaussian wave packet was solved by Nauenberg in \cite{nauenberg2016}. Nauenberg found that a solution \(\psi\) to the Schr{\"o}dinger equation in a frame accelerated with acceleration \(a\) with respect to an inertial frame may be constructed from any solution \(\phi(x,t)\) from the unaccelerated frame as
\begin{equation}
  \label{eq:2}
  \psi(x,t) = \phi \left(x + \frac{at^2}{2}, t \right)
  \exp
  \left[
    - \frac{i mat}{\hbar} \left( x + \frac{at^2}{6} \right)
  \right].
\end{equation}
Choosing \(\phi\) to solve the free Schr{\"o}dinger equation and using Nauenberg's results to determine the corresponding wave packet in a frame accelerated by \(a=g\) gives a result which agrees with the moment result up to overall \(x-\)independent phase.

\subsection{E{\"o}tv{\"o}s parameter}
\label{sec:eotvos}

Gravitational fields arising from matter sources exhibit inhomogeneity. In the Newtonian framework, inhomogeneity corresponds to non-linear potentials. The simplest non-linear potential is the quadratic potential. In a quadratic potential, the second derivative \(d^{2}\Phi/ dx^{2}\) is constant ensuring that the center-of-mass dynamics decouple from the quantum moment dynamics, similarly to the previously considered cases. Due to this similarity we do not immediately construct the equations of motion or present a wave packet solution as before. However, if desired, these can be easily derived. The distinctive features arising in the quadratic case are more relevant in the context of atom interferometry, the topic of Section \ref{sec:propagation-phase}. Discussion of this case is postponed until that section.

Quantum effects begin to influence classical dynamics only when the second derivative \(d^{2}\Phi/ dx^{2}\) becomes dependent on \(x\). However, accurately characterizing the source mass distribution with sufficient resolution to resolve the field structure to this order it is challenging. (A recent experiment by Overstreet et al. in \cite{overstreet2022} presents an intriguing counterexample, where deliberate efforts were made to precisely characterize the source mass distribution.) A simpler test, albeit with less far reaching implications, emerges from the consequence that the weak-field geodesic equation predicts a universal acceleration for all objects regardless of the specific geometry. This observation leads to a class of experiments known as E{\"o}tv{\"o}s experiments. These experiments are designed to constrain the normalized differential acceleration between two objects, expressed as:
\begin{equation}
  \label{eq:eotvos}
  \eta(1,2) = \frac{a_1 - a_2}{\bar{a}}.
\end{equation}
In this definition \(\bar{a} = \frac{a_1+a_2}{2}\) represents the average acceleration, and the quantity \(\eta\) is referred to as the E{\"o}tv{\"o}s parameter.

In general relativity, the weak-field geodesic equation (equation (\ref{eq:weak-field-geodesic})) predicts \(\eta(1,2) = 0\) identically for any two objects even when their masses differ, \(m_1\ne m_2\). In a sense, general relativity is constructed as a geometric theory of gravity with the explicit aim reach of arriving at this conclusion. Modern extensions to the standard model and general relativity typically anticipate some deviation from this classical prediction. Consequently, the parameter \(\eta\) serves as a valuable model-independent framework for quantifying violations of the weak equivalence principle. We are particularly interested in whether quantum effects lead to \(\eta \ne 0\) and, if so, at what level these effects become significant.

Let us rephrase the dynamics in terms of canonical variables (\ref{eq:canonicalVariables}) via the Hamilton function
\begin{equation}
  \label{eq:Heff}
  H\left(x,p, s, p_s \right)
  = \frac{p^2}{2m} + \frac{p_s^2}{2m} + m \Phi_{{\rm eff}}(x,s)
\end{equation}
with the quantum-gravitational potential
\begin{equation}
  \label{eq:effectivePotential}
  \Phi_{{\rm eff}}(x,s) =
  \Phi(x) + \frac{1}{2} \frac{d^2 \Phi}{dx^2} s^2
                           + \frac{1}{8}\left( \frac{\hbar}{m} \right)^2 \frac{1}{s^2}.
\end{equation}
Our canonical effective methods provide direct predictions for quantum corrections to the acceleration of a quantum state in a non-uniform gravitational field. The equations of motion for \(x\) and \(s\) in terms of the gravitational field strength \(g(x)\equiv \Phi'\) are
\begin{eqnarray}
  \label{eq:eomx}
  \ddot{x} &=&  - g - \frac{1}{2}\partial_x^2g\, s^2 \\
  \label{eq:eoms}
  \ddot{s} &= &
              - \partial_x g\, s
              + \left(\frac{\hbar}{m}\right)^2 \frac{1}{4 s^3}.
\end{eqnarray}
We read off the anomalous center of mass acceleration
\begin{equation}
  \label{eq:diffacceleration}
  \left\vert
    \frac{d^2}{dt^2}\langle \hat{x}\rangle - (-g(\langle \hat{x} \rangle))
  \right\vert
  = \frac{1}{2} \partial_x^2g \, \Delta(x^2).
\end{equation}
The center of mass acceleration deviates from the local gravitational
field acceleration when both the width of the state and the gravitational field
strength curvature are non-vanishing. Although this is the outcome expected
from classical tidal forces acting on extended objects in an inhomogeneous
field, our canonical formulation is more general because it provides also the
dynamics of \(s=\sqrt{\Delta(x^2)}\). In particular, the final value of the
anomalous acceleration depends on the value of \(s\) which is plainly
mass-dependent because $m$ appears explicitly in $\ddot{s}$. The origin of the
mass-dependence lies ultimately in the quantum requirement to preserve the
uncertainty product, which is defined for moments of $x$ together with $p$,
rather than $\dot{x}$.

We estimate the E{\"o}tv{\"o}s parameter for a delocalized quantum particle as compared to a more localized particle as
\begin{equation}
  \label{eq:eotvosapp}
  \eta \approx
  g^{-1}
  \left\vert
    \frac{d^2}{dt^2}\langle \hat{x}\rangle - (-g(x))
  \right\vert
  = \frac{1}{2} g^{-1}\partial_x^2g \, \Delta(x^2)
\end{equation}
Parameter values suitable for terrestrial experiments are \(g \approx 10 \,\mathrm{m}/\mathrm{s}^2\) and \(\partial_x^2g \approx 10^{-12}/\mathrm{ms}^2\) however the wave packet width \(\Delta(x^2)\) is not independently well constrained by experiment. Equation (\ref{eq:eotvosapp}) indicates a range of values for \(\eta\) from \(\eta \approx 0.5 \times 10^{-33}\) when the wave packet width is atomic scale (\(s \approx 10^{-10}~\mathrm{m} \)) to \(\eta \approx 0.5 \times 10^{-13}\) when the wave packet width approaches the arm-length of typical interferometers (\(s\approx 1~\mathrm{m}\)). This latter value is within the sensitivity range of proposed atom-interferometers \cite{tino2007,tino2013,trimeche2019, altschul2015} and it is possible that proposed future experiments including km-scale underground tests, and space based atom interferometers could reach these dimensions.

Inverting the above reasoning with the experimental constraints of state-of-the-art atom interferometers which have resolutions of nearly \(10^{-11}g\) \cite{asenbaum2020} requires the wave packet width to remain bounded
\begin{equation}
  \label{eq:12}
  s = \sqrt{\Delta(x^2)} \lesssim 10 \,\mathrm{m}.
\end{equation}

\subsection{Gravitational scattering return time}
\label{sec:returntime}

In \cite{davies2004a,davies2004b} Davies considered the possibility that the quantum dynamics of a particle may allow its time of flight to differ systematically from the classical prediction by travelling beyond the classical turning point into the forbidden region of the gravitational potential. Perhaps surprisingly, Davies found no evidence for tunneling delay. Instead, the particle return time adheres to the classical prediction in gravitational fields which are at most quadratic in position and provided that the particle is measured far from the classical turning point. This result does not challenge the status of the weak equivalence principle for quantum phenomena.
However, stationary state analysis of quantum objects tunneling into the classically forbidden region of a potential gives only limited insight into the dynamical problems encountered, particularly in the context of interferometer experiments.

The equations of motion (\ref{eq:eomx}) and (\ref{eq:eoms}) indicate that for low order gravitational potentials where \(\partial_x^2 g\) vanishes, the classical degrees of freedom decouple from the quantum degrees of freedom. It follows directly then that the measured return time for a wave packet in linear or quadratic potentials is identical to that of a classical point particle in agreement with the stationary state calculations of \cite{davies2004a, davies2004b}.
For higher order potentials \(\partial_x^2 g\) does not vanish and instead
couples the spreading motion of the wave packet to the motion of its center of
mass. This outcome was anticipated by \cite{lammerzahl96} and
\cite{hogan2008}, but neither provided quantitative calculations, which would
be quite challenging if based on wave functions.

Estimating an out-and-back time of flight prediction requires integrating the equations of motion. To do so, we specialize to the case of a Newtonian potential where the effective Hamilton function is
\begin{equation}
  \label{eq:HeffNewtonian}
  H(r,p,s,p_s)
  = \frac{p^2}{2m} + \frac{p_s^2}{2m}
  + \frac{U}{2ms^2} - \frac{GMm}{r} - \frac{GMm}{r^3}s^2.
\end{equation}
It can be seen here that the classical Newtonian potential energy has a power law correction of the form
\begin{equation}
  \label{eq:powerlawPotential}
  V(r) = -\frac{GMm}{r} \left[ 1 + \alpha_N \left(\frac{r_0}{r} \right)^{N-1} \right]
\end{equation}
with \(N = 3\), \(\alpha_3 = 1\), and \(r_0 = s\). Power law modifications of this form have previously been studied. In the context of extensions to the standard model, the case \(N = 3\) can be considered as arising from the simultaneous exchange of two massless pseudoscalar particles \cite{fischbach2001}. A powerlaw correction with \(N=3\) also arises from the model of Randall and Sundrum \cite{randall1999} where non-compact warped extra dimensions with warping scale \(r_0\) are considered. If there are indeed Yukawa-style couplings present, then the finite-width effects that we discuss here could confound their detection.

Choosing an arbitrary length scale equal for example to the earth radius, \(r_c=r_e\) and corresponding time, energy, and momentum scales as
\begin{eqnarray}
  t_c &=& \sqrt{\frac{r_e^3}{GM}} \\
  E_c &=& p_c \frac{r_c}{t_c} = \frac{GMm}{r_e}
\end{eqnarray}
gives the non-dimensional Hamilton function and equations of motion
\begin{equation}
  \label{eq:nonDimensionalH}
H(r,p,s,p_s)
    = \frac{p^2}{2} + \frac{p_s^2}{2}
      + \frac{u}{2s^2} - \frac{1}{r} - \frac{s^2}{r^3}
\end{equation}
\begin{eqnarray}
  \label{eq:nonDimensionalEom}
    \ddot{r} &=& - \frac{1}{r^2} \left( 1 + \frac{3s^2}{r^2} \right) \\
    \ddot{s} &=& \frac{u}{s^2} + \frac{2s}{r^3}  
\end{eqnarray}
Unlike the classical case where all free parameters may be scaled out, here a free parameter remains which depends on the particle mass
\begin{equation}
  \label{eq:10}
  u = \frac{\hbar^2/4}{r_c^2p_c^2} = \frac{\hbar^2/4}{GMm^2r_e}.
\end{equation}
This parameter reflects the minimal uncertainty product compared to the scale
of the problem. The same effect could be had in identifying characteristic
length \((r_c)\) and momentum \((p_c)\) scales and rescaling
\(\frac{1}{2}\hbar \rightarrow
\frac{1}{2}\hbar/(r_cp_c)=\frac{1}{2}\tilde{\hbar}\) such that the canonical
commutation relation is dimensionless. Such a phrasing, however useful, tends
to obscure the mass-dependence of the relation because it implies an
$m$-dependent, non-fundamental $\tilde{\hbar}$.

The uncertainty-product enforcing term becomes important when the wave packet is very narrow compared to \(\sqrt{u}\) in which case the wave packet is forced to expand but for terrestrial experiments, \(u\) is entirely negligible. For a neutron moving in the earth's gravitational field near the mean earth radius the numerical value is \(u \approx 10^{-36}\) while for a 10 gram mass in the same conditions \(u \approx 10^{-86}\). Without this term the resulting trajectory is the one predicted from classical tidal effects only and is not mass-dependent. This aligns with the thinking that existing atom interferometers are essentially classical in their operation. The main perturbation to the center of mass trajectory of a wave packet will come from tidal effects and not quantum effects owing to the uncertainty principle.

\begin{figure}
  \centering
  \includegraphics[width=10cm]{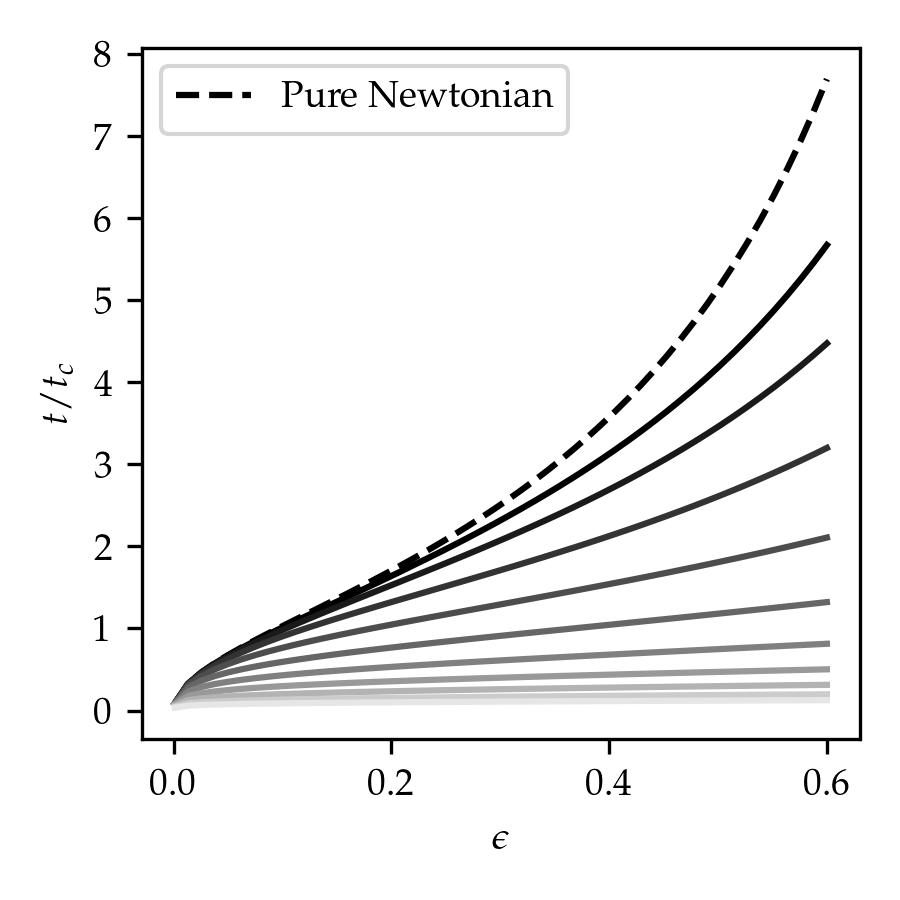}
  \caption{\textbf{Particle return time.} The dashed line plots the out-and-back return time for a classical point particle in Newtonian \(1/r\) gravity. The behavior for small energy is quadratic as predicted from the linear potential result \(t = \sqrt{\epsilon}\). At higher energy the return time grows until it diverges at the classical escape energy \(\epsilon = 1\). The solid lines represent the return time for a particle obeying the perturbed equations of motion (\ref{eq:nonDimensionalEom}) for \(u\in (10^{-5},1)\). These are generically lower than the unperturbed result and approach the classical result as \(u\rightarrow 0\).}
  \label{fig:returnTime}
\end{figure}
Figure \ref{fig:returnTime} presents typical return time curves from numerical integration of the equations of motion for particles whose classical component of the energy is
\begin{equation}
  \label{eq:14}
  \epsilon = \frac{p^2}{2} - \frac{1}{r}.
\end{equation}
The total energy of the particle is still given by equation~(\ref{eq:nonDimensionalH}). In an experiment, the choice of initial conditions depends on the preparation of the state. The ability to discriminate the total energy of the particle from its classical initial conditions may complicate the return time--energy dependence. Here it is evident that a wave packet prepared at a given height and velocity returns quicker than an identically prepared point particle. The effect is most pronounced in the non-linear regime, a result in agreement with our earlier discussion.

\subsection{Propagation phase in interferometry}
\label{sec:propagation-phase}

In 1924, Louis de Broglie introduced a groundbreaking concept through his work \cite{debroglie1924}, suggesting that massive particles possess wave-like characteristics. These wave properties can be understood within the framework of wave functions, where they stem from the polar decomposition described by equation (\ref{eq:polarDecomposition}). A consequence of this description is interference of wave components. A simple example is the interference of two wave function components of equal magnitude where the resulting probability of measuring a particle depends on the relative phase of the two components via
\begin{equation}
  \label{eq:interference}
  \vert \exp(i\theta_1) + \exp(i\theta_2)\vert^2 = 2 + 2 \cos(\theta_1-\theta_2).
\end{equation}
Equation (\ref{eq:interference}) and its dependence on the phase difference \(\delta \theta = \theta_1-\theta_2\) is an example of an interference effect typically ascribed to wave phenomena.

The field of matter wave interference has matured significantly over time and atom interferometer experiments now play a crucial role in a variety of fundamental research endeavors. In this section, we describe how we can gain insights into interference by examining quantum moments. This goal begins with relating quantum moments to the phase difference between spacetime points, denoted \(\theta(x_f,t_f) - \theta(x_i,t_i)\). This quantity is known as the propagation phase of a single wave function component. It is defined rigorously only in the wave function formalism. Nonetheless, we previously described how to determine a position-dependent phase from moments and separately explained the time dependence of these moments. In this section, we bring these concepts together to describe the evolution of the phase along a spacetime trajectory.

As a point of comparison, we first review the approach for calculating the propagation phase presented in \cite{storey1994} based on Feynman path-integral techniques. Other approaches, based on evolving plane waves (\cite{lammerzahl96}) and Gaussian wave packets (\cite{borde2004}) have also been described. These however produce predictions for the phase shift which agree with the semiclassical approach at the level of experiment; see \cite{overstreet2021}.

\subsubsection{Path integral method}
\label{sec:path-integral-method}

In an approach based on the path integral, we consider a state prepared in the wave function state \(\psi(x_i,t_i)\). The quantum evolution of this state is given by the propagator formula
\begin{equation}
  \label{eq:propagatorEvolution}
  \psi(x_f,t_f) = \int dx_i K(x_f,t_f,x_i,t_i) \psi(x_i,t_i).
\end{equation}
The path integral method uses Feynman's expression for the quantum propagator, in which the propagator is represented as a sum over paths \(\Gamma\) connecting the spacetime points \((x_f,t_f)\) and \((x_i,t_i)\) according to
\begin{equation}
  \label{eq:propagatorFeynman}
  K(x_f,t_f,x_i,t_i) = \mathcal{N} \sum_\Gamma \mathrm{e}^{\frac{i}{\hbar} S_{\Gamma}}.
\end{equation}
In this equation, \(S_\Gamma\) represents the action along the path \(\Gamma\).

Storey and Cohen-Tannoudji \cite{storey1994} prove that when the system Lagrangian is quadratic, the quantum propagator (\ref{eq:propagatorFeynman}) can be simplified as
\begin{equation}
  \label{eq:quadraticPropagator}
  K(x_f,t_f,x_i,t_i) = F(t_f,t_i) \exp
  \left(
    \frac{i}{\hbar} S_{{\rm cl}}(x_f,t_f,x_i,t_i)
  \right).
\end{equation}
In this equation \(S_{{\rm cl}}(x_f,t_f,x_i,t_i)\) represents the action along the classical path connecting \((x_f,t_f)\) and \((x_i,t_i)\). When we substitute this expression into equation (\ref{eq:propagatorEvolution}), we obtain:
\begin{equation}
  \label{eq:propagatorEvolution2}
  \psi(x_f,t_f)
  = F(t_f,t_i)
  \int dx_i  \exp
  \left(
    \frac{i}{\hbar} S_{{\rm cl}}(x_f,t_f,x_i,t_i)
  \right)
  \psi(x_i,t_i).
\end{equation}
In the quadratic case, the classical action is a quadratic function of \(x_f\) and \(x_i\).
Therefore, for certain initial states including plane wave and Gaussian states, this integral can be solved in closed form. For a plane wave initial state
\begin{equation}
  \label{eq:3}
  \psi(x_i,t_i) = \frac{1}{\sqrt{2\pi\hbar}} \exp
  \left[
    \frac{i(p_0x_i - E_0t_i)}{\hbar}
  \right].
\end{equation}
In this case, the phase of the integrand is stationary when
\begin{equation}
  \label{eq:stationaryPhaseEqn}
  \frac{\partial S_{{\rm cl}}}{\partial x_i} + p_0 = 0.
\end{equation}
Since \(S_{{\rm cl}}\) is quadratic, equation (\ref{eq:stationaryPhaseEqn}) is
a linear equation for the stationary phase point \(x_{i,{\rm stationary}} \equiv x_0\).
The expansion of the classical action around this point is given by
\begin{equation}
  \label{eq:4}
  S_{{\rm cl}}(x_f,t_f,x_0+ \zeta,t_i)
  =
  S_{{\rm cl}}(x_f,t_f,x_0,t_i)
  -
  p_0 \zeta
  + C(t_f,t_i) \zeta^2.
\end{equation}
In this expression, we used equation (\ref{eq:stationaryPhaseEqn}) to replace
the first derivative of the action with the negative plane wave momentum. We
also introduced the second derivative \(\partial^2 S_{{\rm cl}}/\partial x_i^2 \equiv C(t_f,t_i)\), which is assumed to be independent of \(x_f\) and \(x_i\). Following these adjustments, the integral (\ref{eq:propagatorEvolution2}) becomes straightforward to evaluate, yielding the result:
\begin{equation}
  \label{eq:5}
  \psi(x_f,t_f) = F(t_f,t_i)
  \sqrt{\frac{i\pi\hbar}{C(t_f,t_i)}}
  \psi(x_0,t_i)
  \exp
  \left(
    \frac{i}{\hbar} S_{{\rm cl}}(x_f,t_f,x_0,t_i)
  \right).
\end{equation}
From this expression we read off the propagation phase accumulated between spacetime points as
\begin{equation}
  \label{eq:propagatorPhaseDiff}
  \theta(t_f,x(t_f)) - \theta(t_i,x(t_i))
  = \frac{1}{\hbar}\int_{t_i}^{t_f}L_{{\rm classical}}(x,\dot{x}) dt.
\end{equation}
The classical Lagrangian \(L_{{\rm classical}}(x,\dot{x})\) is treated as a function of time once we specify the classical trajectory \(x(t)\).

\subsubsection{Moment method}
\label{sec:moment-method}

In the moment approach we obtained the reconstruction formula for the phase derivative \(d\theta/dx\). Considering now the phase as a function of space and time, \(\theta = \theta(x,t)\), it is appropriate to consider the reconstruction formula (\ref{eq:generalPhaseDerivativeReconstruction}) as providing the partial derivative with respect to position appearing in the differential
\begin{equation}
  \label{eq:dtheta}
  \mathrm{d} \theta
  = \frac{\partial \theta}{\partial x} \mathrm{d}x + \frac{\partial \theta}{\partial t} \mathrm{d}t.
\end{equation}
Both of the partial derivatives may depend on the coordinates \(x\) and \(t\). For example, the second order Hermite reconstruction depends explicitly on the position coordinate from equation (\ref{eq:secondOrderReconstructionTheta})
\begin{equation}
  \label{eq:20}
  \frac{\partial\theta}{\partial x}
  =
  \frac{\langle \hat p \rangle}{\hbar}
  + (x- \langle \hat x \rangle)
  \frac{ \Delta(xp) }
  {\hbar \Delta(x^2)}\,.  
\end{equation}
With this equations of motion for the moments supplied, this becomes also a function of time, \(\partial\theta/\partial x = \partial\theta/\partial x (x,t)\). If the phase where genuinely a multivariate function, then to determine the phase difference between two spacetime points from differential data would require a line integral of the differential (\ref{eq:dtheta})
\begin{equation}
  \label{eq:21}
  \theta(x_f,t_f) - \theta(x_i,t_i)
  =
  \int_\gamma 
  \left(
    \frac{\partial \theta}{\partial x} \mathrm{d}x + \frac{\partial \theta}{\partial t} \mathrm{d}t
  \right)
\end{equation}
where \(\gamma\) is a path connecting the spacetime points \((x_f,t_f)\) and \((x_i,t_i)\). We can define the path arbitrarily by a parametrization \(\gamma : [\tau_i,\tau_f]\rightarrow \mathbb{R}^2, \gamma(\tau) = (x_\gamma(\tau),t_\gamma(\tau))\) where we require the coordinate functions satisfy \(x_\gamma(\tau_{i/f}) = x_{i/f}\) and \(t_\gamma(\tau_{i/f}) = t_{i/f}\). That is, we compute the parametrized line integral
\begin{equation}
  \label{eq:21exp}
  \theta(x_f,t_f) - \theta(x_i,t_i)
  =
  \int_{\tau_i}^{\tau_f}
  \left(
    \frac{\partial \theta}{\partial x} \left(x_\gamma(\tau),t_\gamma(\tau)\right)
    \frac{\mathrm{d}x_\gamma}{\mathrm{d}\tau} 
    +
    \frac{\partial \theta}{\partial t} \left(x_\gamma(\tau),t_\gamma(\tau)\right)
    \frac{\mathrm{d}t_\gamma}{\mathrm{d}\tau} 
  \right)
  \mathrm{d}\tau
\end{equation}

There are some difficulties in this approach. For one, the moment data does not directly constrain the partial derivative \(\partial \theta / \partial t\). However, if we require that the result obtained be independent of the integration path, then the mixed partial condition
\begin{equation}
  \label{eq:mixedPartials}
  \frac{\partial}{\partial t} \frac{\partial \theta}{\partial x}
  = \frac{\partial}{\partial x} \frac{\partial \theta}{\partial t}
\end{equation}
will allow us to reconstruct \(\partial \theta / \partial t\) from integration of \(\partial \theta / \partial x\) up to an overall time dependent function. For example, when we use the first order result for \(\partial \theta / \partial x\) and substitute the time dependence of moments appropriate for a particle in a linear gravitational field we have
\begin{equation}
  \label{eq:23}
  \frac{\partial\theta}{\partial x}(x,t)
  =
  \frac{\langle \hat p \rangle_0 - m g (t-t_0)}{\hbar}
\end{equation}
which, being the plane wave approximation, is a trivial function of position. Partial differentiating with respect to time gives
\begin{equation}
  \label{eq:23b}
  \frac{\partial}{\partial t} \frac{\partial\theta}{\partial x}
  =
  - \frac{ m g}{\hbar}.
\end{equation}
Using the mixed partial condition (\ref{eq:mixedPartials}) and integrating with respect to position gives the time partial in this case as
\begin{equation}
  \label{eq:23c}
  \frac{\partial\theta}{\partial t}
  =
  - \frac{ m g x}{\hbar} + f(t)
\end{equation}
where the arbitrary function of time \(f(t)\) is added without affecting the mixed partial equality. This residual freedom of time-dependence cannot be eliminated using moment data alone and represents the non-physical arbitrary phase which may be included in any wave function under the scaling (\ref{rescale}). Nonetheless, in this case we are free to choose this arbitrary function of time as
\begin{equation}
  \label{eq:17}
  f(t) = - \frac{\langle \hat{p}\rangle(t)^2}{2m\hbar}
\end{equation}
where \(\langle \hat{p}\rangle(t)\) is the classical time-dependence of the momentum. Putting everything together, we have
\begin{eqnarray}
  \label{eq:21b}
  \theta(x_f,t_f) - \theta(x_i,t_i)
  &=&
    \int_\gamma 
    \left(
    \frac{\partial \theta}{\partial x} \mathrm{d}x
    +
    \frac{\partial \theta}{\partial t} \mathrm{d}t
    \right) \nonumber\\  
  &=&
    \int_\gamma 
    \left[
    \frac{\langle \hat p \rangle(t)}{\hbar} \mathrm{d}x
    -
    \left(
    \frac{\langle \hat{p}\rangle(t)^2}{2m\hbar}
    +
    \frac{ m g x}{\hbar}
    \right) \mathrm{d}t
    \right]
\end{eqnarray}
This line integral was constructed to be independent of choice of integration path. It is convenient to choose the integration path parametrized by time, \(\gamma = (\langle \hat{x} \rangle(t), t)\). Then the line integral is
\begin{equation}
  \label{eq:phaseDifferenceIntegral}
  \theta(x_f,t_f) - \theta(x_i,t_i)
  =
  \frac{1}{\hbar}
  \int_\gamma 
  \left[
  \langle \hat p \rangle
  \frac{\mathrm{d}\langle \hat{x} \rangle}{\mathrm{d}t}
  -
  H
  \right]
  \mathrm{d}t
\end{equation}
where \(H\) is the conserved energy
\begin{equation}
  \label{eq:22}
  H =
  \frac{\langle \hat{p}\rangle^2}{2m}
  +
  m g \langle \hat{x} \rangle.
\end{equation}
The integrand in (\ref{eq:phaseDifferenceIntegral}) is numerically equal to the classical Lagrangian evaluated along the classical trajectory
\begin{equation}
  \label{eq:24}
  \langle \hat p \rangle
  \frac{\mathrm{d}\langle \hat{x} \rangle}{\mathrm{d}t}
  -
  H\left(\langle \hat{x} \rangle, \langle \hat p \rangle\right)
  = L_{{\rm classical}}(t)
\end{equation}
This demonstrates equality between the moment approach and the result derived from the Feynman path integral, expressed in equation (\ref{eq:propagatorPhaseDiff}).

The plane wave approximation determines the propagation phase using only classically-defined quantities which respect the weak equivalence principle in the absence of quantum back-reaction. Quantum back-reaction occurs only if higher-order structure of the gravitational field can be resolved.
Therefore, in low-order gravity-resolving atom interferometer phase measurements, we anticipate no observed violation of the weak equivalence principle. This expectation is supported by experiments, including: (i) a series of simultaneous dual-species atom-interferometer E{\"o}tv{\"o}s tests, presented in \cite{bonnin2013,zhou2015,asenbaum2020}, which constrained \(\eta(^{85}\mathrm{Rb},^{87}\mathrm{Rb})<10^{-12}\); (ii) the dual-species test conducted by \cite{schlippert2014}, which placed constraints on the more significant mass gap, \(\eta(^{39}\mathrm{K},^{87}\mathrm{Rb}) < 10^{-7}\); and (iii) the work of \cite{rosi2017}, which constrained the differential acceleration for atoms in a coherent superposition of metastable energy states at the \(10^{-9}\) level.

These analyses highlight the importance of distinguishing between an atom interferometer's use of quantum properties in making a measurement and the absence of back-reaction of the quantum properties on the measurement. As discussed in \cite{hogan2008} and revisited more recently in \cite{overstreet2021}, the null results obtained so far indicate that quantum back-reaction on the center of mass dynamics either does not manifest, or itself conforms to the equivalence principle. Our analysis suggests that quantum effects do not conform to the equivalence principle. Therefore, it is reasonable to deduce that current instruments lack the detection sensitivity to resolve wave packet effects, i.e. the atoms used in these experiments mimic classical test particles, at least as far as their center of mass motion is concerned.

Our moment expansion, accommodates this conclusion through the moment hierarchy defined in equation (\ref{eq:hierarchy}). However, it also offers a new framework for performing calculations in the regime where quantum back-reactions become important. The challenge of incorporating the wave packet structure intrinsic to atomic test masses prepared for a local experiment has received comparatively little attention, with the work \cite{borde2004} being an exception.

\subsubsection{Incorporating wave packet effects}
\label{sec:incorp-wave-pack}

The potential for quantum back-reaction onto the classical trajectory is highly interesting since it would signal a deviation from the geodesic motion predicted by general relativity, providing a unique regime for testing the compatibility of gravity and quantum mechanics. 
The effect of higher-order potentials on non-local wave packet structure is mass-dependent with the non-zero E{\"o}tv{\"o}s parameter calculated in Section \ref{sec:eotvos}. This quantum effect may be incorporated into the phase determination with the second-order accurate result, equation (\ref{eq:secondOrderReconstructionTheta}):
\begin{equation}
  \label{eq:eq:secondOrderReconstructionThetaCopy}
  \frac{\partial\theta}{\partial x} =
  \frac{\langle \hat p \rangle}{\hbar}
  + (x- \langle \hat x \rangle)
  \frac{ \Delta(xp) }
  {\hbar \Delta(x^2)}.
\end{equation}
An example of the application of this formula could be to identify wave packet effects on the propagation phase.

\begin{figure}
  \centering
  \includegraphics[scale=.75]{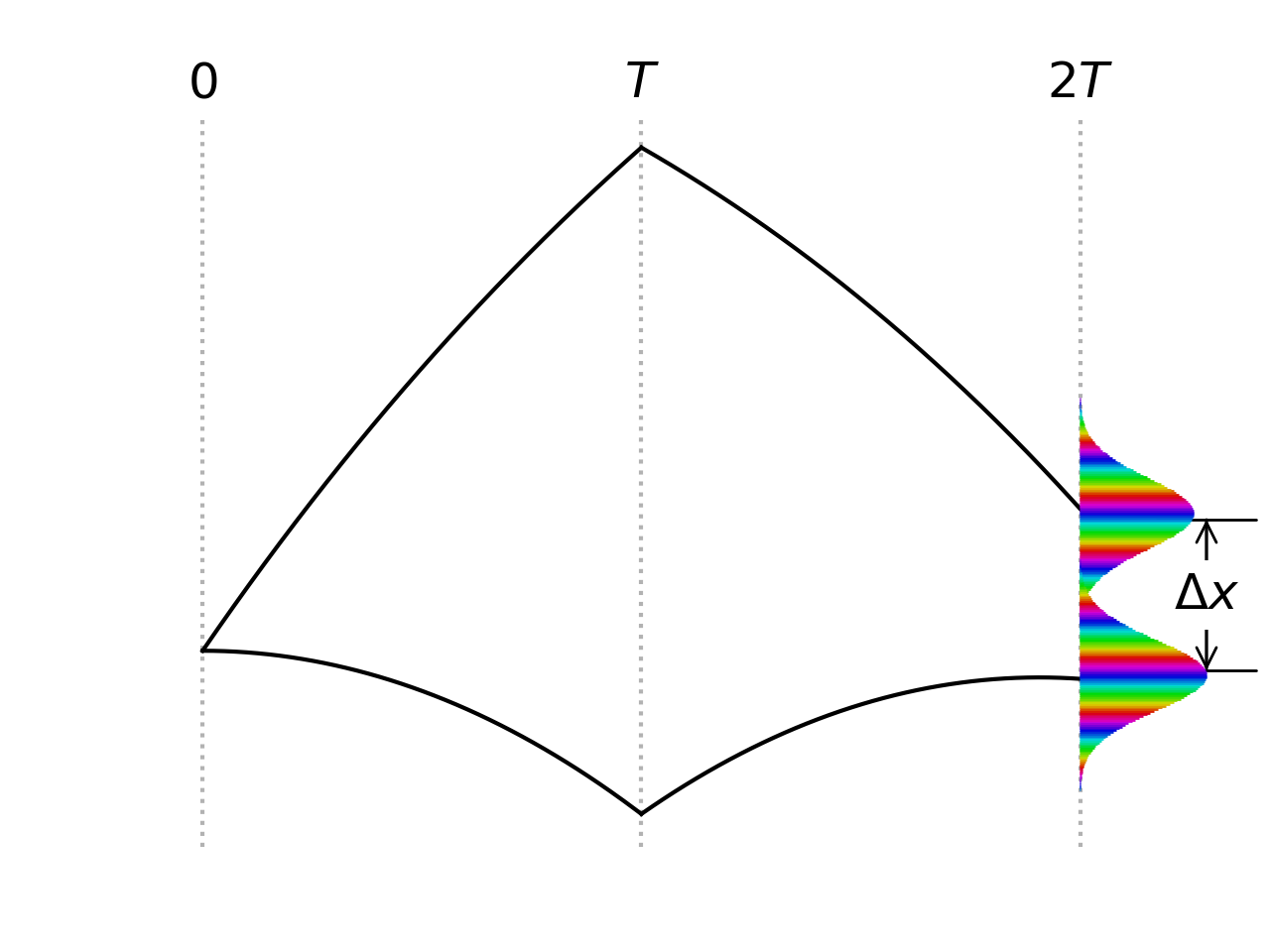}
  \caption{Mach-Zehnder interferometer spacetime geometry in an inhomogeneous gravitational field. The center of mass trajectories of the separated wave packet components do not intersect at the time of the third equally spaced pulse. Instead, spatial separation implies that wave packet structure will determine the interferometric phase difference. The phase of each component is most easily computed by integrating the differential phase along a piecewise path traveling first along the two center of mass trajectories, and then vertically at the fixed end time.}
  \label{fig:MZgeometry}
\end{figure}

We consider a simple Mach-Zehnder interferometer and imagine that the device operates based on light-pulses coherently splitting, redirecting, and recombining an atomic wave packet at equally spaced times. The spacetime geometry is sketched in Figure \ref{fig:MZgeometry}. In this sketch it is made evident that, in the presence of a linear gravity gradient, particle trajectories within a Mach-Zehnder interferometer do not close at the time of an equally spaced pulse after a single reflection.
We can determine the phase of either wave packet component at the time of the recombining pulse, \(t=2T\), at any vertical displacement most easily if we choose an integration path which follows the component's center of mass trajectory until \(t=2T\), and then follows a vertical path at the fixed time \(t=2T\). This choice of integration path is convenient because for the first two segments of the piecewise path the path satisfies \(x_\gamma = \langle \hat{x} \rangle\) and the second-order contributions to the phase vanish. Then in the last segment the moments are all time-independent yielding a simple integration.

Lastly, we comment on the structure of the second-order result. The first order phase difference, adequate for describing plane waves propagating in low-order potentials, was determined by the classical Lagrangian action
\begin{equation}
  \label{eq:7}
  \delta \theta = \frac{1}{\hbar} \int L_{{\rm classical}} dt.
\end{equation}
The moment approach with its canonical structure evaluates the same phase difference, but in phase space coordinates. At the lowest order this was expressed by the Legendre transform of the classical Lagrangian as
\begin{equation}
  \label{eq:7H}
  \delta \theta = \frac{1}{\hbar} \int \left( p\dot{x} - H_{{\rm classical}}\right) dt.
\end{equation}
The second-order formula for the phase derivative in
canonical phase space coordinates (\ref{eq:canonicalVariables}) was
\begin{equation}
  \label{eq:9b}
  \frac{\partial\theta}{\partial x} =
  \frac{ p }{\hbar}
  +
  \frac{x- \langle \hat x \rangle}{s}
  \frac{ p_s }  {\hbar}.
\end{equation}
Integrating over a physical trajectory and changing variables with the Jacobian
\begin{equation}
  \label{eq:11}
  \frac{ds}{dx} = \frac{ x - \langle \hat{x} \rangle }{s}
\end{equation}
yields
\begin{equation}
  \label{eq:9}
  \delta\theta =
  \frac{1}{\hbar} \int
  \left(
    p dx
    +
    p_s ds
  \right).
\end{equation}
The integration is over the phase space trajectory $x(t)$, on which $s(t)$
depends through $s^2=(x-\langle\hat{x}\rangle)^2+{\rm constant}$ according to
(\ref{eq:11}). To this, we may incorporate the time-dependent contribution from
the conserved energy, \(H = \langle \hat{H} \rangle\), for free without
leaving the same Hilbert space ray and finally obtain the
propagation phase
\begin{equation}
  \label{eq:9}
  \delta\theta =
  \frac{1}{\hbar} \int
  \left(
    p \dot{x}
    +
    p_s \dot{s}
    -
    H
  \right) dt
\end{equation}
where \(H= \langle \hat{H} \rangle\) is not equal to the classical Hamilton function but is the effective quantum energy including moments, for example as given by (\ref{eq:Heff}). This final result connects the phase contribution from second-order wave packet structure to the Lagrangian formulation. We expect that this result is obtainable also from the propagator method outlined in \cite{storey1994} because the integrals involved remain Gaussian. Nonetheless, it appears the quantum propagation phase is more naturally understood through its dependence on the quantum phase space structure.

\section{Conclusion}
\label{sec:conclusion}

Although it has been known for some time that several mathematical ingredients
of quantum dynamics are generally mass-dependent even in the case of gravitational forces, there remain questions about the extension of the equivalence principle to quantum mechanics. With this motivation, we set out a simple analysis for the behavior of a quantum object in a (possibly inhomogeneous) gravitational field. In place of stationary state analysis, we considered the physically motivated case of a wave packet following a nearly classical trajectory whose fluctuations remain bounded by powers of \(\hbar\).

The presented moment expansion systematically bridges between classical and
quantum dynamics but at all steps, the method's mathematical description of
the dynamics takes a classical form. The (quasi)classical nature of the
dynamical system permits intuitive dynamical interpretations to our findings
which we have evaluated for the E\"otv\"os parameter of a test mass in free
fall, for the return time of a quantum test mass in a gravitational
field, and for the propagation phase of a quantum object transiting an interferometer. In all cases, quantum fluctuations---or the spreading of a wave packet---imply specific corrections to the classical equations of motion and affect
physical conclusions.

Wave packet spreading is always mass-dependent, even for a free
particle. Whether this mass-dependence noticeably affects the center of mass
motion is a matter of precision. At the classical end, when no quantum
fluctuations are kept, the center of mass motion of a freely falling particle
is independent of mass. When quantum fluctuations are considered, the
mass-dependence of the center of mass motion depends on the functional form of
the potential used, or the order of its Taylor expansion. Owing to the
position of the derivative in Ehrenfest's equation, when the gravitational
potential is at most quadratic in position, the resulting equations of motion
close on the center of mass and are mass-independent in keeping with the
findings of previous studies. However, for higher order potentials (e.g.
Newtonian) we have demonstrated that inhomogeneities in the gravitational
field create quantum tidal forces. The tidal force has the same form here as
in classical calculations, with the addition of a mass-dependent term
enforcing uncertainty constraints on the second-order statistics. In the
coupling of this spreading behavior to the center of mass we find the center
of mass dynamics become mass-dependent as well.

The observed coupling is coarse in that it reflects only the second-order
statistics of the wave packet and then only in the direction of motion. If
further precision is required, then following the logic of
Section~\ref{sec:methods}, this framework may be extended to examine the tidal
effects of higher order fluctuations. In pursuing this one could use the
canonical mappings of higher order fluctuations obtained in \cite{Bosonize} up
to fourth order. Should the full three-dimensional structure of the wave
packet be considered, that reference includes in addition canonical mappings
for more than one degree of freedom. The systematic derivation of these
mappings using methods from Poisson geometry for quantum moments implies
computational advantages compared with a many-body treatment that would be
required for classical tidal effects of mass distributions. If $\hbar$-terms
such as our $u$ are ignored, the quantum derivation may also be interpreted as
a shortcut for a description of the classical effects.

In the case that tidal forces affect the dynamics, the magnitude of the
influence on the center of mass motion may be determined by an E{\"o}tv{\"o}s
parameter $\eta$. However, owing to differences in state preparation across experiments and because the wave packet width in interferometer experiments is typically not independently well-constrained, it is difficult to judge a value for \(\eta \). Values for this effect corresponding to atomic-scale sized wave packets are orders of magnitude below current experimental bounds. The smallness of the mass-dependence for these conditions is a consequence of the smallness of the only free parameter appearing in the dynamics: \(u = \hbar^2/(4GMm^2r_e)\). If the wave packet width is permitted to approach the size of a meter then E{\"o}tv{\"o}s parameter values near the sensitivities reached by existing experiments are plausible to obtain.

In summary, a quantum weak equivalence principle for expectation values is correct only in the limit that tidal effects are irrelevant i.e., the width of the wave packet is small compared to the curvature length of the field. Our analysis uses only non-relativistic quantum mechanics. Such analysis highlights features particular to the non-relativistic theory. We hope our viewpoint is sufficiently clear as to remove any uncertainty which may persist on the topic of universal free fall in quantum mechanics. This framework may prove useful in connection with further tests of the weak equivalence principle.

\section*{Acknowledgements}

This work was supported in part by NSF grant PHY-2206591.


\begin{thebibliography}{10}

\bibitem{carroll}
Sean~M. Carroll,
\newblock An Introduction to General Relativity: Spacetime and Geometry, pages
  153--154,
\newblock Cambridge University Press, New York, NY, 2019

\bibitem{tino2020}
G.M. Tino, L.~Cacciapuoti, S.~Capozziello, G.~Lambiase, and F.~Sorrentino,
\newblock Precision gravity tests and the Einstein Equivalence Principle,
\newblock {\em Progress in Particle and Nuclear Physics} 112 (2020) 103772

\bibitem{greenberger1968}
Daniel {Greenberger},
\newblock The role of equivalence in Quantum Mechanics,
\newblock {\em Annals of Physics} 47 (1968) 116--126

\bibitem{viola1997}
Lorenza Viola and Roberto Onofrio,
\newblock Testing the equivalence principle through freely falling quantum
  objects,
\newblock {\em Phys. Rev. D} 55 (1997) 455--462

\bibitem{sonego1995}
Sebastiano {Sonego},
\newblock {Is there a spacetime geometry?},
\newblock {\em Physics Letters A} 208 (1995) 1--7

\bibitem{okon2011}
Elias Okon and Craig Callender,
\newblock Does quantum mechanics clash with the equivalence principle---and
  does it matter?,
\newblock {\em European Journal for Philosophy of Science} 1 (2011)
  133--145

\bibitem{kasevich1991}
Mark Kasevich and Steven Chu,
\newblock Atomic interferometry using stimulated Raman transitions,
\newblock {\em Phys. Rev. Lett.} 67 (1991) 181--184

\bibitem{kasevich1992}
M.~Kasevich and S.~Chu,
\newblock Measurement of the gravitational acceleration of an atom with a
  light-pulse atom interferometer,
\newblock {\em Applied Physics B} 54 (1992) 321--332

\bibitem{peters2001}
A~Peters, K~Y Chung, and S~Chu,
\newblock High-precision gravity measurements using atom interferometry,
\newblock {\em Metrologia} 38 (2001) 25

\bibitem{hogan2008}
Jason~M. Hogan, David M.~S. Johnson, and Mark~A. Kasevich,
\newblock Light-pulse atom interferometry, 2009

\bibitem{nobili2020}
Anna~M. Nobili, Alberto Anselmi, and Raffaello Pegna,
\newblock Systematic errors in high-precision gravity measurements by
  light-pulse atom interferometry on the ground and in space,
\newblock {\em Phys. Rev. Res.} 2 (2020) 012036

\bibitem{strocchi1966}
F.~Strocchi,
\newblock {Complex coordinates and quantum mechanics},
\newblock {\em Rev.\ Mod.\ Phys.} 38 (1966) 36--40

\bibitem{kibble1979}
T.~W.~B. Kibble,
\newblock {Geometrization of quantum mechanics},
\newblock {\em Communications in Mathematical Physics} 65 (1979) 189 -- 201

\bibitem{ashtekar1999}
Abhay Ashtekar and Troy~A. Schilling,
\newblock Geometrical Formulation of Quantum Mechanics, In {\em On Einstein's
  Path: Essays in Honor of Engelbert Schucking}, pages 23--65,
\newblock Springer New York, New York, NY, 1999

\bibitem{bojowald2006}
Martin Bojowald and Aureliano Skirzewski,
\newblock Effective Equations of Motion for Quantum Systems,
\newblock {\em Rev. Math. Phys.} 18 (2006) 713--745

\bibitem{bjelakovic2005}
Igor Bjelakovi{\'{c}} and Werner Stulpe,
\newblock The Projective Hilbert Space as a Classical Phase Space for
  Nonrelativistic Quantum Dynamics,
\newblock {\em International Journal of Theoretical Physics} 44 (2005)
  2041--2049

\bibitem{bojowald2022}
Martin Bojowald,
\newblock Canonical description of quantum dynamics*,
\newblock {\em Journal of Physics A: Mathematical and Theoretical} 55 (2023) 504006

\bibitem{lammerzahl96}
Claus Lämmerzahl,
\newblock On the equivalence principle in quantum theory,
\newblock {\em General Relativity and Gravitation} 28 (1996) 1043–1070

\bibitem{aragon-munoz2020}
L.~Arag\'{o}n-Mu\~{n}oz, G.~Chac\'{o}n-Acosta, and H.~Hernandez-Hernandez,
\newblock Effective quantum tunneling from a semiclassical momentous approach,
\newblock {\em International Journal of Modern Physics B} 34 (2020) 2050271,
  [https://doi.org/10.1142/S0217979220502719]

\bibitem{Weinstein}
A.~Cannas da~Silva and A.~Weinstein,
\newblock Geometric models for noncommutative algebras, 1999

\bibitem{VariationalEffAc}
R.~Jackiw and A.~Kerman,
\newblock Time-dependent variational principle and the effective action,
\newblock {\em Physics Letters A} 71 (1979) 158--162

\bibitem{GaussianDyn}
F.~Arickx, J.~Broeckhove, W.~Coene, and P.~Van~Leuven,
\newblock Gaussian wave-packet dynamics,
\newblock {\em International Journal of Quantum Chemistry} 30 (1986) 471--481,
  [https://onlinelibrary.wiley.com/doi/pdf/10.1002/qua.560300741]

\bibitem{QHDTunneling}
Oleg~V. Prezhdo,
\newblock Quantized Hamilton Dynamics,
\newblock {\em Theoretical Chemistry Accounts} 116 (2006) 206--218

\bibitem{Bosonize}
Bekir Bayta\c{s}, Martin Bojowald, and Sean Crowe,
\newblock Faithful realizations of semiclassical truncations,
\newblock {\em Annals of Physics} 420 (2020) 168247, [arXiv:1810.12127]

\bibitem{EffPotRealize}
Bekir Bayta\ifmmode~\mbox{\c{s}}\else \c{s}\fi{}, Martin Bojowald, and Sean
  Crowe,
\newblock Effective potentials from semiclassical truncations,
\newblock {\em Phys. Rev. A} 99 (2019) 042114, [arXiv:1811.00505]

\bibitem{brizuela2014}
David Brizuela,
\newblock Classical and quantum behavior of the harmonic and the quartic
  oscillators,
\newblock {\em Phys. Rev. D} 90 (2014) 125018

\bibitem{kjeldsen1993}
Tinne~Hoff Kjeldsen,
\newblock The Early History of the Moment Problem,
\newblock {\em Historia Mathematica} 20 (1993) 19--44

\bibitem{schmudgen2017}
Konrad Schm{\"u}dgen,
\newblock The Complex Moment Problem, In {\em The Moment Problem}, pages
  381--398,
\newblock Springer International Publishing, Cham, 2017

\bibitem{storey1994}
{Pippa Storey} and {Claude Cohen-Tannoudji},
\newblock The Feynman path integral approach to atomic interferometry.
  A~tutorial,
\newblock {\em J. Phys. II France} 4 (1994) 1999--2027

\bibitem{peters1998}
Achim Peters,
\newblock {\em High precision gravity measurements using atom interferometry},
\newblock Phd thesis, Stanford University, Stanford, CA, 1998

\bibitem{nauenberg2016}
Michael Nauenberg,
\newblock Einstein's equivalence principle in quantum mechanics revisited,
\newblock {\em American Journal of Physics} 84 (2016) 879--882,
  [https://doi.org/10.1119/1.4962981]

\bibitem{overstreet2022}
Chris Overstreet, Peter Asenbaum, Joseph Curti, Minjeong Kim, and Mark~A.
  Kasevich,
\newblock Observation of a gravitational Aharonov-Bohm effect,
\newblock {\em Science} 375 (2022) 226--229,
  [https://www.science.org/doi/pdf/10.1126/science.abl7152]

\bibitem{tino2007}
G.M. Tino, L.~Cacciapuoti, K.~Bongs, Ch.J. Bordé, P.~Bouyer, H.~Dittus,
  W.~Ertmer, A.~Görlitz, M.~Inguscio, A.~Landragin, P.~Lemonde, C.~Lammerzahl,
  A.~Peters, E.~Rasel, J.~Reichel, C.~Salomon, S.~Schiller, W.~Schleich,
  K.~Sengstock, U.~Sterr, and M.~Wilkens,
\newblock Atom interferometers and optical atomic clocks: New quantum sensors
  for fundamental physics experiments in space,
\newblock {\em Nuclear Physics B - Proceedings Supplements} 166 (2007)
  159--165,
\newblock Proceedings of the Third International Conference on Particle and
  Fundamental Physics in Space

\bibitem{tino2013}
G.M. Tino, F.~Sorrentino, D.~Aguilera, B.~Battelier, A.~Bertoldi, Q.~Bodart,
  K.~Bongs, P.~Bouyer, C.~Braxmaier, L.~Cacciapuoti, N.~Gaaloul, N.~Gürlebeck,
  M.~Hauth, S.~Herrmann, M.~Krutzik, A.~Kubelka, A.~Landragin, A.~Milke,
  A.~Peters, E.M. Rasel, E.~Rocco, C.~Schubert, T.~Schuldt, K.~Sengstock, and
  A.~Wicht,
\newblock Precision Gravity Tests with Atom Interferometry in Space,
\newblock {\em Nuclear Physics B - Proceedings Supplements} 243-244 (2013)
  203--217,
\newblock Proceedings of the IV International Conference on Particle and
  Fundamental Physics in Space

\bibitem{trimeche2019}
A~Trimeche, B~Battelier, D~Becker, A~Bertoldi, P~Bouyer, C~Braxmaier,
  E~Charron, R~Corgier, M~Cornelius, K~Douch, N~Gaaloul, S~Herrmann, J~Müller,
  E~Rasel, C~Schubert, H~Wu, and F~Pereira dos Santos,
\newblock Concept study and preliminary design of a cold atom interferometer
  for space gravity gradiometry,
\newblock {\em Classical and Quantum Gravity} 36 (oct 2019) 215004

\bibitem{altschul2015}
Brett Altschul, Quentin~G. Bailey, Luc Blanchet, Kai Bongs, Philippe Bouyer,
  Luigi Cacciapuoti, Salvatore Capozziello, Naceur Gaaloul, Domenico Giulini,
  Jonas Hartwig, Luciano Iess, Philippe Jetzer, Arnaud Landragin, Ernst Rasel,
  Serge Reynaud, Stephan Schiller, Christian Schubert, Fiodor Sorrentino, Uwe
  Sterr, Jay~D. Tasson, Guglielmo~M. Tino, Philip Tuckey, and Peter Wolf,
\newblock Quantum tests of the Einstein Equivalence Principle with the
  STE–QUEST space mission,
\newblock {\em Advances in Space Research} 55 (2015) 501--524

\bibitem{asenbaum2020}
Peter Asenbaum, Chris Overstreet, Minjeong Kim, Joseph Curti, and Mark~A.
  Kasevich,
\newblock Atom-Interferometric Test of the Equivalence Principle at the
  ${10}^{\ensuremath{-}12}$ Level,
\newblock {\em Phys. Rev. Lett.} 125 (2020) 191101

\bibitem{davies2004a}
P~C~W Davies,
\newblock Quantum mechanics and the equivalence principle,
\newblock {\em Classical and Quantum Gravity} 21 (2004) 2761

\bibitem{davies2004b}
P~C~W Davies,
\newblock Transit time of a freely falling quantum particle in a background
  gravitational field,
\newblock {\em Classical and Quantum Gravity} 21 (2004) 5677--5683

\bibitem{fischbach2001}
E.~Fischbach, D.~E. Krause, V.~M. Mostepanenko, and M.~Novello,
\newblock New constraints on ultrashort-ranged Yukawa interactions from atomic
  force microscopy,
\newblock {\em Phys. Rev. D} 64 (2001) 075010

\bibitem{randall1999}
Lisa Randall and Raman Sundrum,
\newblock An Alternative to Compactification,
\newblock {\em Phys. Rev. Lett.} 83 (1999) 4690--4693

\bibitem{debroglie1924}
Louis de~Broglie,
\newblock XXXV. A tentative theory of light quanta,
\newblock {\em The London, Edinburgh, and Dublin Philosophical Magazine and
  Journal of Science} 47 (1924) 446--458,
  [https://doi.org/10.1080/14786442408634378]

\bibitem{borde2004}
Christian~J. Bord{\'e},
\newblock Quantum Theory of Atom-Wave Beam Splitters and Application to
  Multidimensional Atomic Gravito-Inertial Sensors,
\newblock {\em General Relativity and Gravitation} 36 (2004) 475--502

\bibitem{overstreet2021}
Chris Overstreet, Peter Asenbaum, and Mark~A. Kasevich,
\newblock Physically significant phase shifts in matter-wave interferometry,
\newblock {\em American Journal of Physics} 89 (2021) 324--332,
  [https://doi.org/10.1119/10.0002638]

\bibitem{bonnin2013}
A.~Bonnin, N.~Zahzam, Y.~Bidel, and A.~Bresson,
\newblock Simultaneous dual-species matter-wave accelerometer,
\newblock {\em Phys. Rev. A} 88 (2013) 043615

\bibitem{zhou2015}
Lin Zhou, Shitong Long, Biao Tang, Xi~Chen, Fen Gao, Wencui Peng, Weitao Duan,
  Jiaqi Zhong, Zongyuan Xiong, Jin Wang, Yuanzhong Zhang, and Mingsheng Zhan,
\newblock Test of Equivalence Principle at $1{0}^{\ensuremath{-}8}$ Level by a
  Dual-Species Double-Diffraction Raman Atom Interferometer,
\newblock {\em Phys. Rev. Lett.} 115 (2015) 013004

\bibitem{schlippert2014}
D.~Schlippert, J.~Hartwig, H.~Albers, L.~L. Richardson, C.~Schubert, A.~Roura,
  W.~P. Schleich, W.~Ertmer, and E.~M. Rasel,
\newblock Quantum Test of the Universality of Free Fall,
\newblock {\em Phys. Rev. Lett.} 112 (2014) 203002

\bibitem{rosi2017}
G~Rosi, G~D'Amico, L~Cacciapuoti, F~Sorrentino, M~Prevedelli, M~Zych, {\v
  C}~Brukner, and G~M Tino,
\newblock Quantum test of the equivalence principle for atoms in coherent
  superposition of internal energy states,
\newblock {\em Nat. Commun.} 8 (2017) 15529

\end{thebibliography}

\end{document}